\documentclass[12pt]{article}
\usepackage{jheppub}

\usepackage{wasysym}
\usepackage{amsmath}
\usepackage{amssymb,amsfonts,amsxtra,mathrsfs,makeidx,graphics,graphicx,amsthm,bbm}
\usepackage{epsfig}
\usepackage[all]{xy}
\usepackage{slashed}
\usepackage{caption}

\newcommand{\CK}{\mathcal{K}}
\newcommand{\CL}{\mathcal{L}}
\newcommand{\CM}{\mathcal{M}}
\newcommand{\CN}{\mathcal{N}}

\newcommand{\CV}{\mathcal{V}}

\renewcommand{\Im}{{\rm Im}}
\renewcommand{\Re}{{\rm Re}}
\newcommand{\Tr}{\mbox{Tr}}

\newcommand{\IR}{\mathbb{R}}

\newcommand{\half}{\frac{1}{2}}
\newcommand{\ndt}{\noindent}

\def\p{\partial}

\def\bea{\begin{eqnarray}}
\def\eea{\end{eqnarray}}
\def\be{\begin{equation}}
\def\ee{\end{equation}}
\def\ba{\begin{align}}
\def\ea{\end{align}}
\def\bse{\begin{subequations}}
\def\ese{\end{subequations}}
\newcommand{\ben}{\begin{eqnarray}}
\newcommand{\een}{\end{eqnarray}}

\newcommand{\bem}{\begin{pmatrix}}
\newcommand{\eem}{\end{pmatrix}}

\def\={\;  = \;}
\def\+{\, + \,}

\def\wt{\widetilde}
\def\wh{\widehat}
\def\bar{\overline}

\def\rt2{\sqrt{2}}

\renewcommand{\Im}{\mbox{Im}}
\renewcommand{\Re}{\mbox{Re}}

\newcommand{\SBH}{S^\text{qu}_\text{BH}}

\def\a{\alpha}

\def\l{{\lambda}}

\def\ve{\varepsilon}

\def\nv{n_\text{v}}
\def\nh{n_\text{h}}

\title{Functional determinants, index theorems, and exact quantum black hole entropy}

\author[1]{Sameer Murthy}
\author[2]{and Valentin Reys}
\affiliation[1]{Department of Mathematics, King's College London \\
The Strand, London WC2R 2LS, U.K.}
\affiliation[2]{Nikhef theory group, Science Park 105, \\
1098 XG Amsterdam, The Netherlands}

\abstract{
The exact quantum entropy of BPS black holes can be evaluated using localization in supergravity. 
An important ingredient in this program, that has been lacking so far, is the one-loop effect
arising from the quadratic fluctuations of the exact deformation (the $Q\CV$ operator). 
We compute the fluctuation determinant for vector multiplets and hyper multiplets around~$Q$-invariant 
off-shell configurations in four-dimensional $\CN = 2$ supergravity with $AdS_{2} \times S^{2}$ boundary 
conditions, using the Atiyah-Bott fixed-point index theorem and a subsequent zeta function regularization. 
Our results extend the large-charge on-shell entropy computations in the literature to a regime of 
finite charges. Based on our results, we present an exact formula for the quantum entropy of BPS black holes 
in~$\CN=2$ supergravity. 
We explain cancellations concerning $\frac18$-BPS black holes in $\CN=8$ supergravity 
that were observed in arXiv:1111.1161. 
We also make comments about the interpretation of a logarithmic term in the topological string partition 
function in the low energy supergravity theory.
}


\begin{document}

\maketitle

\section{Introduction and summary: \\ Quantum entropy of supersymmetric black holes \label{introduction}}

Consider a supersymmetric black hole in a four-dimensional theory of supergravity in asymptotically flat space, 
coupled to~$(\nv+1)$ gauge fields, and carrying electric and magnetic charges~$(q_{I},p^{I})$, $I=0,1,\cdots \nv$, 
under these gauge fields. 
The near-horizon configuration of such a black hole is itself a fully supersymmetric solution 
of the theory, and can be decoupled and studied in its own right as a consistent quantum gravitational system. 
The classical near-horizon field configuration, and the classical entropy of the black hole, are determined in terms of 
the black hole charges, according to the well-known attractor mechanism~\cite{Ferrara:1995h}.

The attractor equations, as presented in~\cite{Ferrara:1995h} for a two-derivative theory of supergravity, 
followed from the BPS equations in the near-horizon region, and the entropy of the black hole was given by the 
Bekenstein-Hawking formula~\cite{Bekenstein:1973ur, Hawking:1974sw}. These ideas were generalized to 
theories including higher-derivative interactions in~\cite{LopesCardoso:1998wt, LopesCardoso:2000qm}, 
by using an off-shell formulation of supergravity, and by using the more general Bekenstein-Hawking-Wald entropy 
formula~\cite{Wald:1993nt, Iyer:1994ys, Jacobson:1994qe}. 
These methods have allowed us to completely understand the BPS black hole entropy for any theory of 
supergravity based on a local effective action.

There is a useful reformulation of the attractor mechanism that relies only on the existence of a 
bosonic~$SL(2) \times SU(2)$ symmetry in the near-horizon region~\cite{Sen:2005wa}. 
This symmetry fixes the value of all the fields up to undetermined constants -- the 
geometry is~$AdS_{2} \times S^{2}$ with overall size~$v$, the gauge fields have a constant electric 
field strength~$e^{I}_*$ on the $AdS_{2}$ factor and a constant magnetic flux on the 2-sphere with charge~$p^{I}$, 
and the scalar fields take constant values~$u^{a}$.
 The classical equations of motion then take the form of the extremization equations for the constant parameters: 
\be \label{Legendre}
\frac{\p \CL^{\rm eff}}{\p v} \=  0 \, , \qquad \frac{\p \CL^{\rm eff}}{\p u^{a}} \=0 \, , \qquad 
\frac{\p \CL^{\rm eff}}{\p e^{I}_*} \= q_{I} \, ,
\ee
where $\CL^{\rm eff}(v,\,e^{I}_*,\,u^{a}, q_{I}, p^{I})$ is the local effective Lagrangian (possibly containing 
higher-derivative interactions) integrated over the~$S^2$ factor and evaluated on the near-horizon configuration. 
The Bekenstein-Hawking-Wald entropy of the black hole is then equal to the Legendre transform of the effective
 Lagrangian~$\CL^{\rm eff}$ at the attractor values of the various fields determined 
by~\eqref{Legendre}:\footnote{The Bekenstein-Hawking-Wald entropy is sometimes referred to as the ``classical'' entropy 
because it relies on a local effective action. We stress that this action can include higher-derivative interactions, 
e.g.~coming from integrating out the massive modes of the theory. Perhaps a better notation would 
be ``Wilsonian entropy'' -- in contrast to the ``exact entropy'', defined in~\eqref{qef}, that we study in this paper.}
\be \label{Sclass}
S^\text{class}_\text{BH}\= -\pi q_{I} \,e^{I}_* - \pi\CL^\text{eff}|_{\text{attr.}} \, . 
\ee
The equations~\eqref{Legendre}, \eqref{Sclass} are a concise and elegant way to recast the classical entropy of 
BPS black holes as a variational principle in the near-horizon region.

To include the effect of quantum fluctuations of the supergravity fields on the BPS black hole entropy, 
\cite{Sen:2008vm} promotes the above variational principle to a functional integral, called the \emph{quantum entropy},
over all the fields of the theory that asymptote to the attractor configuration specified by~\eqref{Legendre}.   
More precisely, it is the expectation value of the Wilson line
\be \label{qef}
\exp \bigl( \SBH (q, p) \bigr) \equiv W (q, p) 
= \left\langle \exp[-i \, q_I \oint_{\tau}  A^I]  \right\rangle_{\rm{AdS}_2}^\text{finite}\ . 
\ee 
The angular brackets indicate an integration (with an appropriate measure) over all the field
fluctuations weighted by the exponential of the Wilsonian effective action at some fundamental scale defining the 
theory such as the string scale, and the superscript denotes a regularization of the divergences that 
arise from the infinite volume of~$AdS_{2}$.

Our goal here is the exact evaluation of this functional integral, for which we use the technique of supersymmetric 
localization applied to supergravity~\cite{Banerjee:2009af, Dabholkar:2010uh, Dabholkar:2011ec, Gupta:2012cy, 
Murthy:2013xpa, Dabholkar:2014ema}, a development that was 
seeded by the powerful application of such methods to supersymmetric field theory~\cite{Pestun:2007rz} 
(see~\cite{Pestun:2014mja, Hosomichi:2015jta} and references therein for very recent reviews). 
As we shall discuss below, the localization technique reduces a complicated quantum functional 
integral to the evaluation of a related integral in the semi-classical limit, i.e.~keeping only 
its classical and one-loop contributions. In the context of $\CN=2$ supergravity coupled to matter multiplets, 
the reduction to a specific semi-classical integral was established in~\cite{Dabholkar:2010uh}, wherein 
the classical part of the computation was performed. 
In this paper, we compute the one-loop determinants of the matter field (vector and 
hyper multiplets) fluctuations. As we shall see, this is an important ingredient in the localization recipe, in 
the absence of which the final result lacks consistency.

The starting point of the localization method to compute a functional integral like~\eqref{qef} 
is the existence of a fermionic symmetry~$Q$ that is realized off-shell in the theory, 
and that squares to a compact~$U(1)$ symmetry. 
One deforms the Lagrangian by a positive-definite $Q$-exact term~$\lambda Q \CV$, with $\lambda \in \IR^{\ge 0}$ 
and $\CV$ an appropriately chosen fermionic functional. 
The exactness guarantees that the functional integral is independent of the 
deformation parameter~$\lambda$. On taking the~$\lambda \to \infty$  limit, 
the problem reduces to a semi-classical evaluation of the original integral over the critical points of
~$Q\CV$.\footnote{A rigorous treatment of the above argument uses the methods of equivariant cohomology, 
and the result is given by the Duistermaat-Heckman-Berline-Vergne-Atiyah-Bott localization formula~\cite{Atiyah:1984px, Duistermaat:1982vw, Berline:1982}. We shall follow the treatment of~\cite{Pestun:2007rz} where 
these methods are nicely explained in field theory language.}

The set of critical points, called the localization locus~$\CM_{Q}$, is a drastically reduced -- 
often finite-dimensional -- space compared to the infinite-dimensional field space that we begin with.  
The choice $\CV =  \int d^{4} x \, \sum_{i} \, \left(Q \psi_{i}\, , \, \psi_{i}\right)(x)$, where the summation runs over 
all fermions~$\psi_{i}$ of the theory and~$\left(. \, , \, .\right)$ is an appropriate positive-definite inner product 
in Euclidean signature, is particularly convenient. For this choice, 
the bosonic localization locus is the set of all solutions of $Q\psi_{i}=0$, i.e.~the zero modes of~$Q$. 
The operator~$Q\CV$ vanishes on this locus, and the final answer consists of an integral 
over the zero modes of~$Q$ of the exponential of the full original action times the quadratic fluctuation 
determinant of the~$Q\CV$ operator around the localization locus.

In the black hole context, we choose a supercharge~$Q$ such that $Q^{2} = L_{0} - J_{0}$, 
where $L_{0}$ is the $U(1)$ rotation of the $AdS_{2}$ and $J_{0}$ is a rotation of the $S^{2}$ 
in the fixed asymptotic~$AdS_{2} \times S^{2}$ region. 
For a theory of~$\CN=2$ supergravity coupled to~$\nv$ physical vector multiplets, the conformal supergravity 
formalism~\cite{deWit:1979ug, deWit:1980tn} provides an off-shell closure of the supersymmetry algebra. 
In this context, the localizing manifold is labelled by~$(\nv+1)$ real parameters~$\{\phi^{I}\}$, $I=0,\cdots, \nv$, 
and the result of localizing the functional integral~\eqref{qef} 
takes the form~\cite{Dabholkar:2010uh}:\footnote{The superscript on the left-hand side indicates that we will consider an all-order perturbation theory result around the leading saddle point. There may be additional non-perturbative contributions, for example from orbifold configurations~\cite{Banerjee:2008ky, Murthy:2009dq, Dabholkar:2014ema}.}
\begin{equation} \label{integral}
 W^\text{pert} (q, p) = \int_{\mathcal{M}_{Q}}  \, \prod_{I=0}^{\nv} d\phi^{I} \, \exp\Big(- \pi  \, q_I  \, \phi^I 
 + 4 \pi \, \Im{F \big((\phi^I+ip^I)/2 \big)} \Big)
 \, Z_\text{det}(\phi^{I})\, ,
\end{equation}
where~$F(X^{I})$ is the holomorphic prepotential of the $\CN=2$ supergravity theory 
(suppressing for now the dependence of~$F$ on the Weyl multiplet, which is taken to be fixed to its attractor value). 
This formula shares a number of interesting features with the OSV proposal~\cite{Ooguri:2004zv}, and 
it is part of the attempt to derive and refine this conjecture from the gravitational theory. 
Details of the comparison with the the original OSV proposal are given in~\cite{Dabholkar:2010uh, Dabholkar:2011ec}. We shall make a comparison with the related proposal of~\cite{Denef:2007vg} in Section~\S\ref{ExFor}.

In this paper we focus on the determinant factor~$Z_\text{det}$ in \eqref{integral} which is the 
main remaining problem in the derivation of the exact gravitational quantum entropy formula. 
This factor~$Z_\text{det}$ includes the measure factor arising from the intrinsic curvature of the 
localization manifold, as well as the 1-loop determinant of quadratic fluctuations of the deformation 
action~$Q\CV$ around the localization manifold:~$Z_\text{det} = Z^\text{ind}_{\text{det}} \, Z_\text{1-loop}$. 
The measure~$Z^\text{ind}_{\text{det}}$ has been discussed (in a slightly different context)
in~\cite{Cardoso:2008fr}.
The task that we set ourselves here is to compute the one-loop fluctuation determinant for 
the~$Q\CV$ operator for vector multiplets and hyper multiplets. The computation of the 
graviton and gravitini determinants is under progress~\cite{deWitMurReys}. 
We compute the determinant of the fluctuations of the fields in the theory normal to the localization 
manifold, at an arbitrary point~$\phi^{I}$, focusing here on the dependence of this determinant on the charges
and on the fields~$\phi^{I}$ and ignoring overall numerical constants.
A non-trivial dependence on~$\phi^{I}$ means that the non-zero modes (under $Q$) of bosons and fermions 
do not cancel in the functional integral. As we will see, the dependence of the determinant on the fields~$\phi^{I}$
appears only through the scale of the fluctuating geometry. 

In the vector multiplet, the gauge-fixing condition does not commute with the off-shell supersymmetry, and 
to treat this problem, we develop a formalism to treat BRST symmetries for vector multiplets 
consistent with the off-shell closure of the supersymmetry algebra. We do so using the standard rules of 
quantization for theories with multiple gauge invariances~\cite{Batalin:1981jr, deWit:1978cd}. 
Our results are applicable to four-dimensional~$\CN=2$ supergravity coupled to vector multiplets in any background 
that preserves some supersymmetry. In the case of the (deformed) 4-sphere, it agrees with the 
treatment of~\cite{Pestun:2007rz, Hama:2012bg}. 
In the~$AdS_{2} \times S^{2}$ background, our formalism leads to a different algebra.  

Our main results concerning black hole entropy are as follows. Firstly, the functional determinants of vector 
and hyper multiplets is given in the concise formula~\eqref{1loopdet}. 
In theories of~$\CN=2$ supergravity, taking the holomorphic prepotential as input, and an assumption
about the induced measure (Equation~\eqref{indmsr}), we derive a perturbatively exact formula for the quantum 
entropy of~$\half$-BPS black holes expressed in Equations~\eqref{Wpert}, \eqref{integral2}. 
Then we explain some non-trivial cancellations 
in theories of extended supergravity that agrees with corresponding microscopic results. 
Finally, we make an observation concerning a term logarithmic 
in one of the localization coordinates in the exact entropy formula. There is a natural interpretation 
of this coordinate as the topological string coupling, thus relating to an existing prediction 
of~\cite{Dabholkar:2005dt, Pioline:2006ni, Denef:2007vg}.

The plan of the paper is as follows. In \S\ref{formalism} we set up the formalism for the calculation of the functional
integral. In \S\ref{ghosts} we deal with the gauge invariance, the BRST cohomology, and the issue of how to combine 
it with the off-shell supersymmetry. In \S\ref{DetCalc} we compute the one-loop determinants of 
the matter multiplets using the zeta-function regularization. In \S\ref{Expansions} we 
discuss large-charge expansions of our results, and compare them to previously obtained results. 
In the final section \S\ref{ExFor}, we present an exact formula for BPS black hole entropy in~$\CN=2$ supergravity,
and comment on the relations to topological strings. 

\vspace{0.1cm}

\noindent \emph{Note added}: While this paper was being prepared for publication, we
received communication from R.~Gupta,~Y. Ito, and I.~Jeon of~\cite{Gupta:2015gga} that contains 
overlapping results.

\section{The set up for the evaluation of quantum entropy \label{formalism}}

In this section we set the stage for the determinant calculations presented in the later sections. 
We first review the formalism of off-shell~$\CN=2$ conformal supergravity in which we work.
We then review BPS black hole solutions in the theory and the corresponding attractor equations. 
Choosing one supercharge~$Q$, we review the localizing equations corresponding to~$Q$, 
and the set of solutions, i.e.~the localizing manifold. We then present the algebra of~$Q$ as 
it acts on the various fluctuating fields of the theory.

\subsection*{The conformal supergravity formalism and the classical black hole \label{confform}}

The $\CN=2$ conformal supergravity~\cite{deWit:1979ug, deWit:1980tn} is a formalism which 
allows for off-shell closure of supersymmetry transformations. The theory describes the 
Weyl multiplet coupled to $(\nv+1)$ vector multiplets labelled by $I = 0, \cdots, \nv$. 
The Weyl multiplet includes the vierbein~$e_\mu^a$, the gravitino fields~$\psi_\mu^i$, 
an antisymmetric tensor~$T_{ab}^{ij}$, as well as other fields needed to close the multiplet off-shell.
The index~$i=1,2$ is a fundamental of the~$SU(2)$ R-symmetry of the theory.
Each vector multiplet contains a gauge field~$A^{I}_\mu$, a complex scalar~$X^{I}$, a real~$SU(2)$ triplet~$Y^I_{ij}$ 
of auxiliary scalars, and the gaugini~$\Omega^I_i$. In this paper, we will only consider abelian vector multiplets. 

The supergravity action that we consider is specified by a holomorphic function called the 
prepotential~$F(X^I,\wh{A})$, describing the coupling of the vector multiplets to the background Weyl 
multiplet through chiral-superspace integrals\footnote{More generally, one can have full-superspace integrals describing 
higher-derivative interactions. It was shown in \cite{Murthy:2013xpa} that a large class of such terms do not 
contribute to the quantum entropy. It would be nice to extend this analysis to the level of a complete proof.}. 
Here,~$X^I$ is the lowest component of the vector multiplet and~$\wh{A} \equiv (T_{\mu\nu}^-)^2$ is the lowest component 
of the chiral multiplet built as the square of the Weyl multiplet. 
This latter dependence encodes higher-derivative terms in the supergravity action proportional to the square 
of the Weyl tensor. Supersymmetry requires that this prepotential be 
holomorphic and homogeneous of degree two,\footnote{The expansion of~$F$ in powers of~$\wh{A}$ stands for 
a derivative expansion in the Lagrangian of the on-shell theory as we discuss  in \S\ref{ExFor} (see~\eqref{FAexp}).}
\be \label{prepXA}
F(\lambda\,X^I,\lambda^2\,\wh{A}) \= \lambda^2\,F(X^I,\wh{A}) \, .
\ee
Electric-magnetic duality of the theory is realized as symplectic transformations under which the 
pair~$(X^{I}, F_{I})$, with~$F_{I} \, \equiv \, \p F(X^I,\hat{A})/\p X^{I}$, transforms linearly. 

The four-dimensional $\CN=2$ superconformal algebra is realized as a local gauge symmetry of this theory. 
As in ordinary gauge theory, one makes a particular choice of gauge in order to perform calculations. 
The physical observables are, of course, gauge invariant.
The superconformal algebra includes a  local dilatation invariance under which the vierbein has scaling 
weight~$w=-1$, and the scalars~$X^{I}$ have~$w=+1$, with associated gauge field~$b_{\mu}$, 
as well as an invariance under special conformal transformations with gauge field~$f_\mu^{\,a}$. 
To gauge-fix the latter, we impose the K-gauge condition~$b_\mu=0$. To gauge-fixing the former, it is convenient 
to introduce the symplectically invariant scalar~$\CK$ via:
\be \label{EminK}
e^{-\CK} \, := \,  -i(X^I \bar{F}_I   - \bar{X}^I F_I) \, . 
\ee 
The field~$e^{-\CK}$ with scaling weight $w=2$ appears in the action as a conformal compensator, with the 
kinetic term for the graviton appearing via the combination:
\be \label{OmLag}
 \sqrt{-g} \,  e^{-\CK} \, R   \, .
\ee
The physical, dilatation-invariant metric is~$G_{\mu\nu} \equiv e^{-\CK} \, g_{\mu\nu}$.

The local scale invariance is generically gauge-fixed by setting a field with non-zero scaling weight to a constant value. 
A common choice of gauge is the condition~$e^{-\CK} =1$ in which we have only~$\nv$ fluctuating vector multiplets. 
In this gauge the original metric~$g_{\mu\nu}$ has the standard Einstein-Hilbert Lagrangian for the graviton, 
as seen easily from the expression~\eqref{OmLag}. 
In this paper we shall use the gauge condition~$\sqrt{-g}=1$ which is also very convenient to analyze our problem~\cite{Dabholkar:2010uh}. 
In this gauge the fluctuations of the graviton~$g_{\mu\nu}$ are 
constrained to have fixed volume, but we gain a linearly acting symplectic symmetry on the $(\nv+1)$ 
freely fluctuating fields~$X^{I}$. 

We see that one of the~($\nv+1$) vector multiplet plays the role of a~\emph{compensating} multiplet. 
In addition, we need another compensating multiplet to gauge-fix the extra gauge symmetries of the conformal 
supergravity theory, and we choose this to be a hyper multiplet as in~\cite{deWit:1980tn}. Unlike the case for vector 
multiplets, a formalism to treat off-shell~$\CN=2$ supersymmetry transformations on hyper multiplets with a finite 
number of auxiliary fields is not known. The compensating hyper multiplet is therefore treated using its equations of 
motion. We will briefly comment on its consequences in the following subsection. 

\vspace{0.4cm}

Conformal~$\CN=2$ supergravity admits a~$\tfrac{1}{2}$-BPS black hole solution with an~$AdS_2\times S^2$ 
near-horizon geometry\footnote{In this paper we only focus on four-dimensional black holes, but the ideas can clearly be 
carried forward to higher-dimensional black holes as well. Steps in this directions have 
been taken in~\cite{Gomes:2013cca}.}. 
The near-horizon solution is fully BPS, as discussed in the introduction. In the gauge~$\sqrt{-g}=1$ chosen above, it has the following form (with all other fields not related by symmetries set to zero):
\bea \label{metric}
&&ds^2 = \left[-(r^2-1)dt^2 + \frac{dr^2}{r^2-1}\right] + \left[d\psi^2 + \textnormal{sin}^2(\psi)d\phi^2\right] \, , \\
\label{fieldconfattr}
&&\;\;\;F^{I}_{rt} = e^I_* \, , \quad {F}^I_{\psi\phi} = p^I\sin\psi \, , \quad X^I = X^I_* \, , \quad T_{rt}^- = w \, .
\eea
Here~$F^I_{\mu\nu}$ is the field strength of the~$U(1)$ vector field in the vector multiplet~$I$,~$(e^I_*,p^I)$ are real constants and~$(X^I_*,w)$ are complex constants.

The full-BPS solution \eqref{metric} has a~$SL(2) \times SU(2)$ bosonic symmetry, the two factors acting on the~$AdS_{2}$ and~$S^{2}$ parts respectively. It also admits eight supersymmetries, which together with the bosonic symmetries form the~$SU(1,1 | 2)$ superalgebra. One of the supercharges that we shall call~$Q$ will play an important role in the following. It obeys the algebra
\be \label{specificQ}
Q^{2} \= L_{0} - J_{0} \, , 
\ee
where~$L_{0}$ and~$J_{0}$ are the~Cartan generators of the~$SL(2)$ and the~$SU(2)$ algebras respectively. 

The attractor equations following from full supersymmetry of the near-horizon geometry (or equivalently using the entropy function formalism) are:
\be \label{attraceq}
e^I_* - ip^I - \frac{w}{2}\bar{X}^I_* = 0 \, , \qquad 4i(\bar{w}^{-1}\bar{F}_I - w^{-1}F_I) = q_I \, , \qquad |w|^2 = 16 \, .
\ee
The phase of the complex number~$w$ parametrizing the near-horizon geometry can be set to zero 
using the~$U(1)_R$ gauge symmetry of the theory, which implies~$w=4$. This choice also fixes the 
value of the field~$\hat{A} = (T^-_{\mu\nu})^2$ to~$\hat{A}=-64$. 
With this choice, the attractor equations for the scalars are:
\be
X^I_* + \bar{X}^I_* = e^I_* \, \qquad X^I_* - \bar{X}^I_* = ip^I \, , 
\ee
and  
\be
F_I\big((e_{*}^I+ip^I)/2\big) - \bar{F}_I\big((e_{*}^I-ip^I)/2\big)\big|_{\hat{A}=-64}  = iq_I.
\ee
For such a black hole, using~\eqref{Sclass}, the attractor entropy is~\cite{Sahoo:2006rp}:
\be \label{Sattr}
S_{\text{BH}}^{\text{class}} \= - \pi  \, q_I  \, e_{*}^I  + 4 \pi \, \Im\,{F \big((e_{*}^I+ip^I)/2 \big)}\big|_{\hat{A}=-64} \, .
\ee
At the two-derivative level in the supergravity action, 
one may recast the above entropy formula in terms of the field~$\CK$ introduced 
in~\eqref{EminK} as follows~\cite{Ferrara:1995ih}:
\be
S_{\text{BH}}^{\text{class}} \= \pi e^{-\CK} \, .
\ee
In this form, it is clear that if we scale all charges as~$(q_{I},p^{I}) \to \Lambda (q_{I},p^{I})$ with~$\Lambda \to \infty$, 
the classical entropy of the black hole scales as~$\Lambda^2$. We will refer to this scaling behavior later in this paper.

\subsection*{The localization manifold \label{secloc}}

In order to apply localization, we must first Wick-rotate the metric and field configuration to Euclidean signature,
which is implemented via~$t\rightarrow i\tau$ in the metric~\eqref{metric} and the field configuration~\eqref{fieldconfattr}. All  
spinors are four-dimensional symplectic Majorana-Weyl spinors~\cite{VanProeyen:1999ni}. 
In the conformal supergravity, we have the usual $Q$-supersymmetry transformations and 
an additional conformal supersymmetry (called $S$-supersymmetry). 
These transformations are parameterized by the spinors~$\xi^i_\pm$ and~$\eta^i_\pm$, respectively. 
The index~$i=1,2$ is an~$SU(2)$ index and~$\pm$ denotes the chirality of the spinor. 
Our conventions are given in Appendix~\ref{euclideanspinors}.

The BPS equations of conformal~$\CN=2$ supergravity are obtained by requiring that the variations of all
the fermions in the theory vanish. The vanishing variations of the Weyl multiplet fermions
yield the following equations (the details of these equations, including the definitions of the covariant derivative 
are given in Appendix~\ref{AppKilling}):
\be \label{CKS1}
2D_\mu\xi^i_\pm \pm \tfrac{1}{16}T_{ab}^\mp \gamma^{ab}\gamma_\mu\xi^i_\mp - 
\gamma_\mu\eta^i_\mp \= 0 \, ,
\ee
\be \label{CKS2}
\gamma^\mu D_\mu T^\mp_{ab} \gamma^{ab}\xi^i_\mp \pm 24 D\xi^i_\pm - T^\mp_{ab}\gamma^{ab}\eta^i_\pm \= 0 \, .
\ee
These equations are known as conformal Killing spinor equations in the literature.  
The field~$D$ that appears in~\eqref{CKS2} is an auxiliary scalar field sitting in the Weyl multiplet. In contrast to~\eqref{CKS1}, which determines the Killing spinors of the space-time and thus contains geometrical information, Equation~\eqref{CKS2} does not impose any additional constraints on the geometry and is used to fix the value of the background auxiliary fields~$T_{ab}$ and~$D$ compatible with the conformal Killing spinors.
To apply localization, the first step is to find all bosonic backgrounds that admit spinors~$\xi^{i}_{\pm}, \eta^{i}_{\pm}$ 
obeying the off-shell BPS equations~\eqref{CKS1}, \eqref{CKS2}. 
This problem was analyzed in~\cite{Gupta:2012cy} by using the equation of motion of 
the field~$D$ at the two-derivative level. Note that the equation of motion can of course change upon including 
higher-derivative terms~\cite{LopesCardoso:2000qm}. This problem remains to be analyzed with an appropriate off-shell 
treatment of hyper multiplets. Moreover, it was also assumed in~\cite{Gupta:2012cy} that the $SU(2)_R$ gauge field 
remains flat on the localization manifold. It is possible that this expectation be confirmed once the gauge field 
couples to hyper multiplets, but this analysis is beyond our present scope and will not be carried out.
The additional~\emph{on-shell} input gives a relation between the spinors~$\eta^i_{\pm}$ and~$\xi^i_{\pm}$, which, in the 
gauge~$e^{-\CK}=1$, is simply~$\eta^i_\pm=0$. This makes it clear that the conformal Killing spinor equations reduce
to the usual Killing spinor equations\footnote{See~\cite{Klare:2013dka} for an analysis of the full off-shell Euclidean conformal Killing spinor equations.} (generalized to include the $T_{ab}$ auxiliary field of the Weyl multiplet). 
 
With this condition, one can solve the off-shell BPS equations~\eqref{CKS1} with the attractor  
boundary conditions. The result of~\cite{Gupta:2012cy} is that, in the gauge~$\sqrt{-g}=1$, the only solution 
to these equations\footnote{\label{rsym}This is true modulo the assumption regarding the $SU(2)_R$ gauge field
mentioned above.} is
~$AdS_{2} \times S^{2}$. We present the Euclidean metric in a coordinate system that 
will be useful in the following: 
\be \label{metric2}
ds^2 \= \sinh^2\eta \,  d\tau^2   + d\eta^2 + d\psi^2 + \sin^2\psi \, d\phi^2 \, .
\ee
The coordinate~$\eta$ is related to the coordinate~$r$ in~\eqref{metric} as~$r = \cosh \eta$. 

To find the complete localization manifold, we have to analyze the off-shell BPS equations~$Q\psi_{i}=0$ 
in the vector multiplets as well. These were analysed in~\cite{Dabholkar:2010uh, Gupta:2012cy}, and the result is that 
the solution set is labelled by one real parameter~$C^{I}$ in each vector multiplet:
\be \label{scalars}
X^I (\eta) = X^I_* +  \frac{C^{I}}{\cosh \eta}, \qquad \bar{X}^I (\eta) = \bar{X}^I_* + \frac{C^{I}}{\cosh \eta} \, , 
\qquad Y_1^{I,1} (\eta) = -Y_2^{I,2} (\eta) = \frac{2\, C^{I}}{\cosh^{2}\eta} \, .
\ee
These scalar field fluctuations actually preserve half of the supersymmetries, they do not obey the equations 
of motion, and they are supported by the auxiliary fields~$Y_{ij}^I$ in the vector multiplets. 

The final result is that the full localization manifold of the Weyl multiplet coupled to vector multiplets
is given by~\eqref{metric2}, \eqref{scalars}, thus leading to an~$(\nv+1)-$dimensional localization 
manifold~$\CM_{Q}$. The coordinates on~$\CM_{Q}$ used in the formula~\eqref{integral} are related to the off-shell 
fluctuations in~\eqref{scalars} as:
\be
\phi^{I}=e^{I}_{*}+2\,C^{I} = X^I(0) + \bar{X}^I(0).
\ee

\subsection*{Off-shell supersymmetry transformations and algebra \label{offshellsusy}}

We now move to the supersymmetry transformations of the fluctuations around the localizing manifold. 
The off-shell algebra of our chosen supercharge~$Q$ is given to us by the conformal~$\CN=2$ supergravity 
formalism -- we simply restrict the full off-shell algebra of eight local supercharges to the one supercharge~$Q$
that we focus on.

\vspace{0.2cm}

\noindent \textbf{Vector multiplets:} 
The supersymmetry transformation rules for the vector multiplet using the Killing 
spinor~$\xi^{i}_{(1)}$ given in~\eqref{killingspinor} on our localizing background are (from now on, we drop the 
subscript~$(1)$ on the Killing spinor):
\bea \label{SUSYvect}
QA^I_\mu &=& \epsilon_{ij}\left(\bar{\xi}^i_-\gamma_\mu\,\lambda^{I\;j}_+ - \bar{\xi}^i_+\gamma_\mu\,\lambda^{I\;j}_-\right)\, , \cr
QX^I &=& \epsilon_{ij}\,\bar{\xi}^i_+\lambda^{I\;j}_+ \, , \qquad Q\bar{X}^I = \epsilon_{ij}\,\bar{\xi}^i_-\lambda^{I\;j}_-\, ,  \cr
Q\lambda_+^{I\;i} &=& \tfrac{1}{2}\mathcal{F}^{-\;I}_{ab}\gamma^{ab}\xi_+^i + 2\gamma^\mu\partial_\mu X^I\xi_-^i - Y^{I\;i}_j\xi^j_+ \, , \\
Q\lambda_-^{I\;i} &=& \tfrac{1}{2}\mathcal{F}^{+\;I}_{ab}\gamma^{ab}\xi_-^i - 2\gamma^\mu\partial_\mu\bar{X}^I\xi_+^i - Y^{I\;i}_j\xi^j_- \, , \cr
QY^I_{ij} &=& 2\epsilon_{ki}\epsilon_{jl}\bar{\xi}^{(k}_+\gamma^\mu D_\mu\lambda^{I\;l)}_- - 2\epsilon_{k(i}\epsilon_{j)l}\bar{\xi}^k_-\gamma^\mu D_\mu\lambda^{I\;l}_+\, , \nonumber
\eea
where
\be
\mathcal{F}^{-\;I}_{ab} \equiv F_{ab}^{-\;I} - \frac{1}{4}\bar{X}^I T_{ab}^- \, , \qquad 
\mathcal{F}^{+\;I}_{ab} \equiv F_{ab}^{+\;I} - \frac{1}{4}X^I T_{ab}^+ \, , \nonumber 
\ee
and~$F^{\pm\;I}_{ab}$ is the (anti)self-dual part of the abelian vector field strength~$F^I_{\mu\nu} = 2\partial_{[\mu}A^I_{\nu]}$. 
The covariant derivative acting on spinors is given by~$D_\mu = \partial_\mu - \tfrac{1}{4}\omega_\mu^{\;\;ab}\gamma_{ab}$.

The square of the supersymmetry transformations can be obtained by evaluating 
the full off-shell algebra~\cite{deWit:1979ug, deWit:1980tn} on our 
localizing background (or simply by acting twice with~\eqref{SUSYvect}):
\bea \label{Qsquarevect}
Q^2A^I_\mu &=& i v^\nu F^I_{\nu\mu} + \partial_\mu\left(2K_+\bar{X}^I + 2K_-X^I\right) \, , \cr
Q^2X^I &=& i v^\mu\partial_\mu X^I \, , \qquad Q^2\bar{X}^I = i v^\mu\partial_\mu \bar{X}^I \, ,\\
Q^2\lambda^{I\;i}_+ &=& i v^\mu D_\mu\lambda^{I\;i}_+ + \tfrac{i}{4}D_av_b\gamma^{ab}\lambda^{I\;i}_+ \, , \cr
Q^2\lambda^{I\;i}_- &=& i v^\mu D_\mu\lambda^{I\;i}_- + \tfrac{i}{4}D_av_b\gamma^{ab}\lambda^{I\;i}_- \, , \cr
Q^2Y^I_{ij} &=& i v^\mu \partial_\mu Y^I_{ij} \nonumber \, ,
\eea

The transformation parameters in~\eqref{Qsquarevect} are given by
\be \label{Killingcontract}
v^\mu = -2i\epsilon_{ij}\bar{\xi}^i_+ \gamma^\mu\,\xi^j_-\, , \qquad  \!\!\!\!K_\pm = \epsilon_{ij}\,\bar{\xi}_\pm^i \xi_\pm^j\, . 
\ee
In the right-hand side of \eqref{Qsquarevect}, we use the following useful identities 
\be \label{useid}
iD_{[a}v_{b]} = -\frac{1}{4}K_-T^-_{ab} - \frac{1}{4}K_+T^+_{ab} \, ,
\ee
and 
\be
\partial_\mu K_\pm = \frac{i}{8}v^\nu T_{\mu\nu}^\mp \, , 
\ee 
which can be derived directly from the definition of the Killing vector and the conformal Killing 
spinor equations~\eqref{CKS1} with~$\eta^i_\pm = 0$. 

Using the explicit form of the Killing spinor~\eqref{killingspinor}, we find that 
\be \label{vmu}
v^\mu \=  \bigl( -1 \quad  0 \quad  0 \quad  1 \bigr)^{T} \, , 
\ee
and 
\be \label{kpm}
K_\pm \= \frac{1}{2}\left(\pm\cos\psi - \cosh\eta\right) \, ,
\ee
which we will use in the next section. 

\vspace{0.4cm}

\noindent \textbf{Hyper multiplets:}
We consider a set of~$n_H$ hyper multiplets where the scalars are denoted 
by~$A_i^{\;\;\alpha}$ with~$\alpha=1\ldots 2\,n_H$. The index~$i$ is a doublet under the~$SU(2)$ R-symmetry,
so that we have total of~$4\,n_{H}$ real scalars. 
The~$4\,n_H$ fermions are the~$2\,n_H$ positive-chirality spinors~$\zeta^\alpha_{+}$ and the~$2\,n_H$ 
negative-chirality spinors~$\zeta^\alpha_{-}$. 
We take the hyper multiplet fields to be neutral under the~$U(1)$ gauge symmetry of the vector multiplet, as this is 
consistent with the classical attractor solution in asymptotically flat space. 
The scalars~$A_i^{\;\;\alpha}$ span a quaternionic-K\"{a}hler manifold and we will assume that the 
target-space of the hyper multiplet sigma model is flat~\cite{deWit:1984px}.

Hyper multiplets do not participate in the classical attractor black hole background discussed in~\S\ref{confform}
-- they take zero or constant values as shown in~\eqref{metric}, and as a consequence, they do not contribute to the 
classical action. Their \emph{quantum fluctuations}, however, 
are relevant for our discussion, and we will need an off-shell supersymmetry algebra to treat these fluctuations
within our approach. 
For vector multiplets we could directly use the formalism of off-shell conformal supergravity. For hyper multiplets,  
there is no known off-shell formalism for the full~$\CN=2$ supersymmetry algebra with a finite number of auxiliary 
fields. 

There is, however, a formalism for the off-shell closure of the algebra of one supercharge for vector and 
hyper multiplets with a finite number of auxiliary fields~\cite{Berkovits:1993hx}. This formalism was used in 
localization problems in four-dimensional field theory as in~\cite{Pestun:2007rz, Hama:2012bg}. 
This algebra acting on vector multiplets is exactly the one given by the conformal~$\CN=2$ supergravity
formalism that we used in the previous section. 
As was emphasized in~\cite{Dabholkar:2010uh}, the localization solutions~\eqref{scalars} are \emph{universal} 
in the sense that they do not depend on the physical action of the theory and continue to hold even in the presence 
of other matter fields (which are all constant as in the classical background~\eqref{metric}).\footnote{We shall not 
concern ourselves here with any potentially new solutions to the localization equations in the other matter multiplet sectors. 
The investigations of~\cite{Dabholkar:2011ec} and those below indicate that any such solutions will not contribute to the 
functional integral~\eqref{qef}, but we cannot prove this at the moment.}
We can therefore use the formalism of~\cite{Berkovits:1993hx} and~\cite{Pestun:2007rz, Hama:2012bg} 
for hyper multiplets in black hole backgrounds.

The Q-supersymmetry transformation rules are:
\bea \label{SUSYhyp}
QA_i^{\;\;\alpha} &=& 2\epsilon_{ij}\left(\bar{\xi}_-^j\zeta^\alpha_- - \bar{\xi}^j_+\zeta^\alpha_+\right)\, , \cr
Q\zeta^\alpha_+ &=& \gamma^\mu \partial_\mu A_i^{\;\;\alpha}\xi^i_- + 2\epsilon_{ij}\breve{\xi}^i_+H^{j\,\alpha} \, , \cr
Q\zeta^\alpha_- &=& \gamma^\mu \partial_\mu A_i^{\;\;\alpha}\xi^i_+ + 2\epsilon_{ij}\breve{\xi}^i_-H^{j\,\alpha} \, , \\
QH^{i\,\alpha} &=& \bar{\breve{\xi}}^{\;i}_-\gamma^\mu D_\mu\zeta^\alpha_+ - \bar{\breve{\xi}}^{\;i}_+\gamma^\mu D_\mu\zeta^\alpha_-\, , \nonumber
\eea
where the action of the covariant derivative on the spinors is exactly as in the vector multiplet. 
Here,~$H^{i\,\alpha}$ are~$4\,n_H$ scalar auxilary fields. Indeed, upon setting~$H^{i\,\a}=0$, one recovers the 
on-shell transformation rules of~\cite{deWit:1984px}. 

In the off-shell transformations~\eqref{SUSYhyp}, the parameters~$\breve{\xi}^{\;i}_\pm$ are 
built to satisfy:
\bea \label{constrainedparam}
\bar{\xi}^i_-\breve{\xi}^j_- &=& \bar{\xi}^i_+\breve{\xi}^j_+ \, , \cr
\epsilon_{ij}\bar{\breve{\xi}}^{\;i}_\mp\breve{\xi}^j_\mp &=& \epsilon_{ij}\bar{\xi}^i_\pm\xi^j_\pm \, , \\
\epsilon_{ij}\bar{\breve{\xi}}^{\;i}_+\gamma^\mu\breve{\xi}^j_- &=& \epsilon_{ij}\bar{\xi}^i_+\gamma^\mu\xi^j_- \, . \nonumber
\eea
In these equations, the spinors~$\xi^i_\pm$ are given by~\eqref{killingspinor} as before. 
As mentioned in~\cite{Hama:2012bg}, the constraints~\eqref{constrainedparam} do admit 
non-trivial solutions to~$\breve{\xi}^j_\pm$, and we present an explicit solution for our background in Appendix~\ref{symbol}. 
With these constraints, the Q-supersymmetry transformations close off-shell:
\bea \label{Qsquarehyp}
Q^2A_i^{\;\;\alpha} &=& i v^\mu \partial_\mu A_i^{\;\;\alpha} \, , \cr
Q^2\zeta^\alpha_+ &=& i v^\mu D_\mu\zeta^\alpha_+ + \tfrac{i}{4}D_av_b\gamma^{ab}\zeta^\alpha_+ \, , \\
Q^2\zeta^\alpha_- &=& i v^\mu D_\mu\zeta^\alpha_- + \tfrac{i}{4}D_av_b\gamma^{ab}\zeta^\alpha_-  \, , \cr
Q^2H^{i\,\alpha} &=& i v^\mu \partial_\mu H^{i\,\alpha} \nonumber \, . 
\eea

For the localization analysis, we set all the fermion variations under~$Q$ in~\eqref{SUSYhyp} to zero. 
It is clear that the configuration where the auxiliary field~$H^{i\a}=0$ and the hyper multiplet scalars~$A_{i}^{\a} = \textnormal{constant}$ 
is a solution to the above BPS equations. In order to find an exhaustive list of all solutions, one needs to do an analysis as in~\cite{Gupta:2012cy} 
by separating the different tensor structures on the right-hand side. For now, we proceed with the trivial solutions.

\vspace{0.4cm}

\noindent \textbf{Supersymmetry algebra of $Q$} \\
Inspection of~\eqref{Qsquarevect} and~\eqref{Qsquarehyp} shows that supersymmetry 
algebra of~$Q$ acting on all fields of the vector and hyper multiplets in the~$AdS_{2}\times S^{2}$  background is:
\be \label{Qsquarefullcov} 
Q^2 = i \, \delta_\text{cgct}(v) + i \, \delta_M\left(L_{ab}\right) + \delta_{\text{gauge}}(\theta^I) \, , 
\ee
where the quantities on the right-hand side are as follows. 
The operator~$\delta_\text{cgct}(v)$ is the covariant general coordinate transformation, defined in e.g.~\cite{deWit:1980tn}, 
which is the variation under all gauge symmetries of the conformal supergravity theory (including 
regular general coordinate transformations, but also e.g.~the~$U(1)$ gauge 
symmetry of the vector multiplets), with the gauge parameters determined by the 
vector~$v^\mu$ (given by~\eqref{Killingcontract} for our background). 
In our case, it is equal to the sum of the Lie derivative along the vector~$v^\mu$ and 
the~$U(1)$ gauge transformation parametrized by~$-v^{\mu} A^{I}_{\mu}$. 
The transformation~$\delta_M$ is a Lorentz transformation parametrized by
(see \eqref{useid})
\be
L_{ab} \, := \, \frac{i}{4}\left(K_+T^+_{ab} + K_-T^-_{ab}\right) \= D_{[a}v_{b]} \, ,
\ee
which, on our background solution, equals
\be  \label{ourlab}
L_{ab} \= \begin{pmatrix} 0 & \cosh \eta & 0 & 0 \\ -\cosh \eta & 0 & 0 & 0 \\ 0 & 0 & 0 & \cos \psi \\ 0 & 0 & -\cos \psi 
& 0 \end{pmatrix}\, . 
\ee
Lastly, the transformation~$\delta_{\text{gauge}}$ is a U(1) gauge transformation parametrized by
\be
\theta^I \, := \, 2K_+\bar{X}^I + 2K_-X^I \, .
\ee

In the following, we will combine the off-shell supersymmetry~$Q$ with the BRST symmetry 
encoding the~$U(1)$ gauge symmetry of the vector multiplet. To do so, we isolate the~$U(1)$ gauge 
connection term present in the covariant general coordinate transformation of~\eqref{Qsquarefullcov} and combine it 
with the gauge transformation already present in the algebra of~$Q$.
We thus rewrite the off-shell supersymmetry algebra as\footnote{We note here that a similar procedure 
can be used to combine the spin-connection term appearing in the covariant general coordinate transformation of fermions 
with the Lorentz transformation parameter~$L_{ab}$. In the Lorentz gauge where~$\omega_\tau^{\;12} = -\cosh\eta\, , \; 
\omega_\phi^{\;34} = \cos\psi$, this yields~$\wh{L}_{ab} := L_{ab} - v^\mu\omega_\mu^{\;ab} = 0$, so that the supersymmetry 
algebra is simply~$Q^2 = iv^\mu\partial_\mu + \delta_{\text{gauge}}(\wh{\theta}^I)$. In this paper, we will stay in a generic 
Lorentz gauge where such cancellations need not happen.}
\be \label{Qsquare}
Q^2 \=  i \, \mathcal{L}_v + i \, \delta_M(L_{ab}) + \delta_{\text{gauge}}(\wh{\theta}^I) \, ,
\ee 
where~$\mathcal{L}_v$ is the Lie derivative along the vector~$v$, and
\be
\wh{\theta}^I \, := \, 2K_+\bar{X}^I + 2K_-X^I -i v^\mu A_\mu^I \, .
\ee
Using the values~\eqref{fieldconfattr} of the background gauge fields~$A_{\mu}^{I}$ on the localizing manifold, 
we obtain the explicit expression:
\be \label{ourthhat}
\wh{\theta}^I \= -e^{I}_{*} - 2 C^{I} \= - \phi^{I} \, . 
\ee
Note that the gauge parameters on the right-hand side of the supersymmetry algebra are precisely the coordinates 
on the localizing manifold. 

\vspace{0.2cm}

We note that the algebra~\eqref{Qsquare} of the supercharge~$Q$ is similar in structure, but not quite the same,
as the one appearing in~\cite{Pestun:2007rz,Hama:2012bg}. Before specifying the background manifold, the 
off-shell supersymmetry transformations \eqref{SUSYvect}, \eqref{SUSYhyp} are the same as the corresponding 
ones in~\cite{Pestun:2007rz,Hama:2012bg}.  
The reason for the difference is simply that the background values of all the supergravity fields are different. 
In particular, the right-hand side of the algebra~\eqref{Qsquare} 
involves the~$SU(2)$ R-symmetry of supergravity in the case of the sphere, while this term is absent 
in our case. Instead, the~$AdS_{2} \times S^{2}$ algebra contains a Lorentz rotation which the sphere 
algebra does not have. This fact will play a role in our analysis of the index theorem in~\S\ref{DetCalc}.

\section{Gauge-fixing and the introduction of ghosts \label{ghosts}}

We now turn to the issue of gauge-fixing the~$U(1)$ symmetry in each vector multiplet. 
The main problem is that the action of fixing a gauge does not commute with the off-shell supersymmetry -- 
which is central to our localization methods. 
To treat this problem, we will need to extend the off-shell supersymmetry algebra of~$Q$ to include the 
effect of the gauge-fixing. We also saw a hint of this appearing in the fact that the supercharge~$Q$ squares 
to a compact bosonic generator only modulo a gauge transformation in Equation~\eqref{Qsquare}.

It is natural to solve this problem by combining the conformal~$\CN=2$ supergravity 
formalism with the covariant BRST formalism\footnote{Another, more hands-on method is to choose a 
suitable gauge-fixed background and to compute the bosonic and fermionic eigenmodes around this background. 
The non-cancellation then happens because the naive~$Q$ operator, upon acting on a 
certain eigenmode, moves us out of the gauge-fixing condition and one therefore has to modify~$Q$ as 
in e.g.~\cite{Alday:2013lba}.} by adding Fadeev-Popov ghosts to the theory. 
The technical task is to set up a BRST complex for the gauge symmetries of the theory, and combine it
with the off-shell supersymmetry complex generated by~$Q$. This procedure builds a new supercharge~$\wh{Q}$
which, as we will demonstrate, is suitable for localization and encodes both the gauge symmetry and the 
supersymmetry of the action.
Once this formalism has been set up, the approach turns out to be extremely compact, 
and we can use index theory to elegantly compute the required functional determinants as laid 
out in~\cite{Pestun:2007rz}.

To treat the~$U(1)$ gauge symmetry of the vector multiplet, we introduce a standard BRST ghost system. 
A~$U(1)$ gauge transformation acts on the vector fields as
\be
\delta_g A^I_\mu = \partial_\mu\lambda^I
\ee
where~$\lambda^I$ is the parameter of the transformation in each vector multiplet. To each of these transformations 
we associate a ghost~$c^I$ along with an anti-ghost~$b^I$ and a Lagrange multiplier~$B^I$. Notice that the
operator~$\partial_{\mu}$ has normalizable zero modes on a compact space, namely any constant function. In order 
to treat these zero modes we need to introduce the so-called ghost-for-ghosts: the constant field~$c_0^I$, along with 
two BRST-trivial pairs~$(\bar{\eta}^I,B^I_1)$ and~$(\eta^I,\bar{B}^I_1)$. 
This is the required field content to properly fix the gauge in the path integral~\eqref{integral}. 
This fact is most easily understood by making use of the Batalin-Vilkovisky 
formalism~\cite{Batalin:1981jr, Gomis:1994he} and noting that the gauge theory at hand is 
a first stage reducible theory. 

The BRST transformation laws of the vector multiplet fields in the adjoint of the~$U(1)$ gauge group are:
\be \label{BRSTvect}
\delta_BA^I_\mu   =  \Lambda\,\partial_\mu c^I \, ,  \quad 
\delta_BX^I = 0 \, , \quad  \delta_B\bar{X}^I =0 \, , \quad
\delta_B\lambda^{i\;I}_+  =  0 \, , \quad \delta_B\lambda^{i\;I}_- = 0 \, , \quad
\delta_BY^I_{ij} = 0 \, ,  
\ee
with~$\Lambda$ a constant anti-commuting parameter parametrizing the BRST transformation. We also have the following 
transformations on the ghost fields:
\be
\delta_Bb^I = \Lambda B^I \, , \quad \delta_BB^I = 0 \, , \quad \delta_B\eta^I = \Lambda\bar{B}^I_1 \, , \quad \delta_B\bar{B}^I_1 = 0 \, , \quad \delta_B\bar{\eta}^I = \Lambda B_1^I \, , \quad \delta_BB_1^I = 0 \, ,
\ee 
and
\be
\delta_Bc^I = \Lambda c^I_0 \, , \quad \delta_B c^I_0 = 0.
\ee
The operator~$Q_B$ defined by~$\delta_B\phi := \Lambda\,Q_B\phi$ ($\phi$ being any field of the theory) is a nilpotent operator, due to the fact that the field~$c^I_0$ is constant. 

We now add to the~$\CN=2$ supergravity Lagrangian a~$Q_B$-exact gauge-fixing term:
\be \label{gflagrangian}
\CL_{\text{g.f.}} = Q_B\left[b^I\left(-\frac{B^I}{2\xi_A} + G^A(A^I_\mu)\right) + \bar{\eta}^I\left(-\frac{B^I_1}{2\xi_c} + G^c(c^I)\right) + \eta^I\left(-\frac{\bar{B}^I_1}{2\xi_b} + G^b(b^I)\right)\right] \, , 
\ee
where~$G^A, G^c$ and~$G^b$ are appropriate gauge-fixing functions for the vector field, the ghost and the anti-ghost, respectively, 
and~$\xi_A$,~$\xi_b$ and~$\xi_c$ are constant parameters. The gauge-fixed action
\be
S_{\text{gauge-fixed}} = S_0 + \int d^4x \,\CL_{\text{g.f.}} \, ,
\ee
where~$S_0$ is the action of vector and hyper multiplets coupled to conformal supergravity, is BRST invariant 
since~$\CL_{\text{g.f.}}$ is~$Q_B$-exact and~$Q_B$ is nilpotent. Expanding~\eqref{gflagrangian} using the BRST
transformation rules leads to the expression
\bea \label{gfaction}
S_{\text{g.f.}} &=& \int d^4x \,\CL_{\text{g.f.}} \cr
&=& \int d^4x \,\Big\{B^I\left(G^A(A^I_\mu) - \frac{B^I}{2\xi_A} - \eta^I\frac{\delta G^b}{\delta b^I}\right) - b^I\frac{\delta G^A(A^I_\mu)}{\delta A^J_\mu}\partial_\mu c^J \cr
&&\;\;\;+~\bar{B}^I_1\left(G^b(b^I) - \frac{\bar{B}^I_1}{2\xi_b}\right) + B_1^I\left(G^c(c^I) - \frac{B_1^I}{2\xi_c}\right) - c^I_0\bar{\eta}^J\frac{\delta G^c(c^I)}{\delta c^J}\Big\} \, . 
\eea 

One can recognize in this action the field~$B^I$ as a Gaussian-weighted Lagrange multiplier for the gauge 
condition~$G^A(A^I_\mu) = \eta^J\frac{\delta G^b(b^I)}{\delta b^J}$, the field~$B_1^I$ as a Gaussian-weighted Lagrange multiplier 
for the gauge condition~$G^c(c^I) = 0$ and the field~$\bar{B}_1^I$ as a Gaussian-weighted Lagrange multiplier for the 
gauge condition~$G^b(b^I) = 0$. For the case at hand, these last two gauge-fixing functions are supposed to freeze 
the freedom one has in shifting the ghost and anti-ghost by a constant function, and we can thus take them specifically
to be~$G^c(c^I) = c^I$ and~$G^b(b^I) = b^I$. The~$B_1^I, \bar{B}_1^I$ Lagrange multipliers then impose the conditions that~$\int c^I = 0$ 
and~$\int b^I = 0$, respectively. The gauge-fixing function for the gauge field~$A_\mu^I$ is then fixed to~$G^A(A^I_\mu) = \eta^I$ 
through the equation of motion for the Lagrange multiplier~$B^I$. Note also that the partition function computed from this 
gauge-fixed action is independent of the~$\xi_A,\xi_c$ and~$\xi_b$ parameters~\cite{Pestun:2007rz}.

We pause here for a moment in order to make a technical comment on the ghost set up that was used in~\cite{Pestun:2007rz}. 
For non-abelian gauge theories, like the one considered in~\cite{Pestun:2007rz}, 
constant functions like~$c_0$ are not zero modes of the operator~$D^a_\mu$ (where~$a$ is a color index). 
One could have tried to set up the ghost-for-ghost~$c_0$ to be a zero mode of the covariant 
derivative and thus take it to be a covariantly constant function -- indeed, this may seem natural from a certain
point of view. Doing so, however, would render the integrations over the gauge field and the ghost-for-ghost 
inter-dependent inside the path-integral, which is difficult to implement in practice. 
The strategy for non-abelian gauge fields considered in~\cite{Pestun:2007rz} was to keep~$c_0$ as a constant function, 
and use a BRST charge which is non-nilpotent. In our case the gauge symmetry is abelian.


We now apply the above formalism to our problem of abelian vector multiplets on~$AdS_{2} \times S^{2}$.
The non-compact nature of the space introduces some subtleties. Firstly, we need to specify boundary conditions on all the fields. 
For the physical fields, we choose boundary conditions as in~\cite{Castro:2008ms,Murthy:2009dq}. 
For the ghost fields, we impose Dirichlet boundary conditions on the fields~$b^I,\,c^I$. This implies that 
there is no normalizable zero modes for these fields, and therefore no ghost-for-ghosts. This is consistent 
with the boundary conditions used in~\cite{Sen:2011ba} for the gauge parameters. 
Using this formalism, we set all the ghost-for-ghost fields to zero hereafter.  

Secondly, there is the issue of boundary modes which are normalizable modes of the gauge fields~$A_\mu^I$ that are 
formally pure gauge, but with gauge parameters that do not vanish at infinity
(these have been called ``discrete modes''~\cite{Sen:2011ba}). 
The four-dimensional bulk action depends only on gauge invariant quantities and therefore does not 
depend on these discrete modes -- thus naively giving a divergence in the path integral. 
These special modes have been treated carefully in~\cite{Sen:2011ba}, and the idea is to obtain their contribution 
separately using arguments of ultra-locality. This gives rise to a factor of~$\ell^{-2\beta}$ to the 
functional integral, where $\ell$ is the background length scale of the problem and $\beta$ depends on the field under consideration.
The non-zero modes can be treated as usual, but since we need a complete set of local fields in the computation, 
we should add and subtract one set of zero modes\footnote{In order to justify this procedure more carefully in our localization computation, 
one needs to analyze the cut-off theory and carefully take an infinite-volume limit. This must be done in such a way as to keep
the local superalgebra and the completeness of the basis intact. Another possible resolution of this subtlety is that 
boundary effects will lift these zero modes in the localization action, as consistent with the fact that~$H$ takes non-zero values 
on these modes.
The boundary conditions introduced in the context 
of the $AdS/CFT$ in~\cite{Henningson:1998cd} may be relevant to this discussion.}
to the non-zero modes, thus 
obtaining the contribution of a complete local set of modes and a factor of~$\ell^{2}$.
As a result, we need to multiply the answer found by using a complete set of local field observables by a 
factor~$\ell^{2-2\beta}$.
For the gauge fields, one has~$\beta=1$~\cite{Sen:2011ba},
which effectively means that the discrete modes do not contribute to the determinant calculation.\footnote{In contrast, 
these modes are expected to play a role in the graviton calculation.} 

\vspace{0.4cm} 

\noindent \textbf{The combined supercharge~$\wh{Q}$ and its algebra}\\ 
We now consider the combined transformation for the BRST symmetry and the 
off-shell supersymmetry, generated by~$\wh{Q} \equiv Q+Q_B$. We require this new supercharge to square to
\be \label{Qalg}
\wh{Q}^2 =  i \, \mathcal{L}_v + i \, \delta_M\left(L_{ab}\right) \, \equiv \, H \, , 
\ee
where~$\mathcal{L}_v$ and~$\delta_M$ are the Lie derivative and the Lorentz transformations 
defined around Equation~\eqref{Qsquare}. 
Note that the vector multiplet gauge transformation is no longer present on the right-hand side of the 
algebra~\eqref{Qalg} -- precisely because it is already encoded in the BRST symmetry. 
The above algebra~\eqref{Qalg} allows us to systematically derive the supersymmetry transformation rules 
on the ghost system. 
Expanding~$\wh{Q}^2$, and using the algebra~\eqref{Qsquare} for~$Q$ and the nilpotency of~$Q_{B}$, we obtain
\be
\wh{Q}^2 = Q^2 + Q_B^2 + \left\{Q,Q_B\right\} =  i \, \mathcal{L}_v + i \, \delta_M\left(L_{ab}\right) + \delta_{\text{gauge}}(\wh{\theta}^I) + \left\{Q,Q_B\right\} \, .
\ee
Comparing with~\eqref{Qalg}, we deduce that the anticommutator of a supersymmetry 
and a BRST transformation on the physical and auxiliary fields of the theory should compensate for 
the gauge transformation parametrized by the vector and scalar fields of the vector multiplet. 
Applying this observation to the various fields leads to the supersymmetry transformation rules for 
the ghost system. 

As an example, consider the vector field~$A_\mu^I$:
\be
\left\{Q,Q_B\right\}A_\mu^I \= Q\left(\partial_\mu c^I\right) \= -\partial_\mu(\wh{\theta}^I) \, , 
\ee
which immediately yields 
\be
Qc^I = -\wh{\theta}^I \, .
\ee
Applying~$\wh{Q}^2$ to the other fields of the theory, we obtain the remaining supersymmetry 
transformations\footnote{The same procedure can be applied to also determine the transformation rules for the ghost-for-ghost fields
when they are present, e.g.~as in~\cite{Pestun:2007rz}.}
\be
Qb^I = 0 \, , \quad QB^I = i\mathcal{L}_v b^I \, . 
\ee

We can now write down the various anticommutators on all fields of the theory as
\bea
Q^2\Phi^{(')} &=& \left(i\mathcal{L}_v + i\delta_M(L_{ab})+ \delta_{\text{gauge}}(\wh{\theta}^I)\right)\Phi^{(')}\, , \!\!\qquad Q^2(\text{gh.}) = 0 \, , \cr
Q_B^2\Phi^{(')} &=& 0 \, , \qquad \qquad \qquad \qquad \qquad \qquad \qquad \quad \;\; Q_B^2(\text{gh.}) = 0 \, , \\
\left\{Q,Q_B\right\}\Phi^{(')} &=& -\delta_{\text{gauge}}(\wh{\theta}^I)\Phi^{(')} \, , \qquad \qquad \qquad \quad \; \left\{Q,Q_B\right\}(\text{gh.}) = i\mathcal{L}_v(\text{gh.})\, , \nonumber
\eea
where~$\Phi^{(')}$ stands for bosonic (fermionic) physical and auxiliary fields, and~$\text{gh.}$ stands 
for all the ghost field of the gauge-fixing complex. Using these transformation rules, we conclude that
the complete set of fields (including the ghosts) now admits a symmetry~$\wh{Q}$ realized off-shell with 
algebra~\eqref{Qalg}. This is the supercharge that we would like to use to perform localization, and the localizing
arguments need to be reapplied with this new operator.

The first observation to be made is that the complete gauge-fixed action is closed under~$\wh{Q}$,
\be
\wh{Q}\left(S_0 + S_{\text{g.f.}}\right) = 0 \, .
\ee
This is the case since the~$S_0$ action is gauge and supersymmetry invariant by definition, and as was 
established in~\cite{Pestun:2007rz}, one may replace~$Q_B$ in~\eqref{gflagrangian} by~$\wh{Q}$ without 
changing the value of the path integral under consideration. Thus, the gauge-fixed action we built by 
introducing the gauge-fixing complex is closed under the~$\wh{Q}$ operator, and this operator squares to 
a sum of bosonic symmetries. This is the correct setup for localization. 

We also need to revisit the conditions for the saddle point around which the localization is performed. This means we now look for
solutions to the equation
\be
\wh{Q}\psi_i = Q\psi_i + Q_B\psi_i = 0
\ee
for all physical fermions~$\psi_i$ in the theory. For the gaugini in the adjoint representation of the gauge group, the added 
term~$Q_B\lambda^{I\;i}_\pm$ is zero and therefore does not modify the initial solution found for~$Q\lambda=0$ in~\cite{Dabholkar:2010uh}. 
A similar statement can be made for the fermions of the hyper multiplets. 

%

Finally, we need to modify the deformation operator~$Q\CV$ used in localization to the operator~$\wh{Q}\wh{\CV}$ which now includes 
the gauge-fixing part of the action~\eqref{gflagrangian}:
\be
\wh{\CV} \equiv \CV + \CV_{\text{g.f.}} =  \int d^{4} x \, \left[\sum_{i} \, \left(Q {\psi}_{i}\, , \, \psi_{i}\right) + b^I G^A(A^I_\mu) \right] \, ,
\ee
where, following the discussion below Equation~\eqref{gfaction}, we have discarded the ghost-for-ghost fields and taken the parameter~$\xi_A$ to infinity 
in the gauge-fixing action.
Here we point out that the Euclidean analytic continuation of the spinors that we chose in section 2 is not compatible with 
the positive-definiteness of the action~$\wh{Q}\wh{\CV}$. So, one has to make a choice between supersymmetry and positive-definiteness. 
We choose to preserve supersymmetry, and as we see in the next section, we obtain a sensible final result. We take this to mean 
that for the unpaired modes under $(-1)^F$ (that is, for the index computation), the fluctuation determinant is well-defined. 
The other choice of analytic continuation includes its own complications (e.g. new localizing solutions), as discussed in~\cite{Gupta:2015gga}.  

We now have the full formalism in place to compute the super-determinant of the~$\wh Q \wh \CV$ operator over the 
$\wh{Q}$-complex~\eqref{SUSYvect},~\eqref{BRSTvect},~\eqref{SUSYhyp}, which we proceed to do.

\section{Calculation of the one-loop determinant \label{DetCalc}}

In this section we compute the one-loop determinant of the~$\wh{Q}\wh{\CV}$ operator using an index theorem.
We follow the procedure as explained in~\cite{Pestun:2007rz,Hama:2012bg, LeeSJ, Hosomichi:2015jta}\footnote{We thank Sungjay Lee 
for many informative discussions about this topic.}. 
We will first organize the various fields on which the~$\wh{Q}$ operator acts in 
bosonic and fermionic quantities as:
\be
X^a \xrightarrow{\wh{Q}} \wh{Q} X^a \, ,   \qquad \Psi^\alpha \xrightarrow{\wh{Q}} \wh{Q}\Psi^\alpha\, ,
\ee
where~$X^a$ and~$\Psi^\alpha$ stand for fundamental bosons and fermions, respectively. 
The full set of bosonic and fermionic fields of the theory are thus organized as: 
\be
\mathfrak{B} \, \equiv \, \{X^a\,,\,\wh{Q}\Psi^\alpha\} \; \, (\text{bosonic}) \, , \qquad 
\mathfrak{F}  \, \equiv \, \{\Psi^\alpha\,, \,\wh{Q}X^a\}  \; \, (\text{fermionic})\, .
\ee
The field-splittings for the vector and hyper multiplets are shown in Appendix~\ref{symbol}. 
With this change of variables, the deformation operator~$\wh\CV = \CV + \CV_\text{gf}$ can be written, up to quadratic order in the fields, 
as follows:
\be \label{deformationoperator}
\wh \CV|_{\text{quad.}} \= \left( \wh Q X \; \Psi \right) \; \left( \begin{matrix} D_{00} & D_{01} \\ D_{10} & D_{11} \end{matrix} \right)
\left( \begin{matrix} X  \\ \wh Q \Psi \end{matrix} \right) \, . 
\ee
This implies the following form for~$\wh Q \wh \CV$:
\be  \label{QVLbLf}
\wh Q \wh \CV|_{\text{quad.}} \= \int \, d^{4}x \, \Bigl( \mathfrak{B}  \, K_{b} \, \mathfrak{B}   
\+  \mathfrak{F} \, K_{f} \, \mathfrak{F}  \, \Bigr) \equiv \, \CL_{b} \+ \CL_{f}\, ,   
\ee
\be \label{Lb}
\CL_{b} \= \left(  X \; \wh Q \Psi \right) \; \left( \begin{matrix} H & 0 \\ 0 & 1 \end{matrix} \right) 
\left( \begin{matrix} D_{00} & D_{01} \\ D_{10} & D_{11} \end{matrix} \right)
\left( \begin{matrix} X  \\ \wh Q \Psi \end{matrix} \right) \, , 
\ee
and
\be  \label{Lf}
\CL_{f} \= \left( \wh Q X \; \Psi \right) \; \left( \begin{matrix} D_{00} & D_{01} \\ D_{10} & D_{11} \end{matrix} \right)
\left( \begin{matrix} 1 & 0 \\ 0 & H \end{matrix} \right)
\left( \begin{matrix} \wh Q X  \\  \Psi \end{matrix} \right) \, ,
\ee
and where~$H = \wh Q^{2}$ as defined in~\eqref{Qalg}. 

By definition, the one-loop determinant for the operator~$\wh Q \wh \CV$ is:
\be \label{Z1loop}
Z_\text{1-loop} \= \left(\frac{\det K_{f}}{\det K_{b}} \right)^{\half} \, . 
\ee
From equations \eqref{QVLbLf}, \eqref{Lb}, \eqref{Lf}, we have that
\be \label{detratio}
\frac{\det K_{f}}{\det K_{b}} \= \frac{\det_{\Psi} H}{\det_{X} H} \= 
\frac{\det_\text{Coker$D_{10}$} H}{\det_\text{Ker$D_{10}$} H} \, . 
\ee
The above ratio of determinants can be computed from the knowledge of the index
\be \label{indD10}
\text{ind} (D_{10})(t) := \Tr_\text{Ker$D_{10}$}  \, e^{-iHt} - \Tr_\text{Coker$D_{10}$}  \, e^{-iHt} \, . 
\ee
Indeed the expansion of the index 
\be
\text{ind} (D_{10})(t) \= \sum_{n} a(n) \, e^{-i \lambda_{n} t} \, , 
\ee 
encodes the eigenvalues~$\l_{n}$ of~$H$, as well as their indexed degeneracies~$a(n)$, and 
we can thus write the ratio of determinants in \eqref{detratio} as:
\be \label{detratio1}
\frac{\det_\text{Coker$D_{10}$} H}{\det_\text{Ker$D_{10}$} H} \= \prod_{n} \, \l_{n}^{-a(n)} \, . 
\ee
This infinite product is a formal expression, and we will discuss a suitable regulator in the following. 

From a mathematical point of view, the index~\eqref{indD10} is an \emph{equivariant index} with respect 
to the action of~$H$. This operator acts on all the fields as~$H=i\mathcal{L}_v + i \delta_M(L_{ab})$ 
according to~\eqref{Qalg}. 
The action of~$H$ on the spacetime manifold is simply through the Lie derivative, 
i.e.~the $U(1)$ action~$H= (-i\partial_\tau + i\partial_\phi) \equiv L_{0} - J_{0}$. 
A $U(1)$-equivariant index of this type can be computed in an elegant manner using the Atiyah-Bott 
index theorem for transversally elliptic operators~\cite{Atiyah:1974}, as 
explained in detail in~\cite{Pestun:2007rz}. Here we will make use of this index theorem even though 
$AdS_2$ is a non-compact space. We note in this context that the $AdS$ space is effectively compact, in the sense that there is a 
gravitational potential well that localizes physical excitations around the fixed point of the $U(1)$ action. This 
suggests that continuous modes do not contribute to the index, which is what we will assume. 
We leave a detailed analysis of the boundary conditions and boundary action as an interesting problem to be analyzed 
in the future. We summarize the ideas of the index theorem very briefly from a working point of view
in Appendix~\ref{symbol}, where we also show that  the~$D_{10}$ operator is transversally elliptic with respect to the 
action of~$H$. 
The result of the theorem applied to our problem is that the index of the~$D_{10}$ operator~\eqref{indD10} reduces 
to the fixed points of the manifold under the action of~$H$. 
Denoting this action by $x \mapsto \wt x = e^{-iHt} x$, we have:  
\be \label{ASindthm}
\text{ind} (D_{10})(t) \= \sum_{\{x \mid \wt x = x\}} \frac{\Tr_{X,\Psi} \, (-1)^{F} \, e^{-iHt }}{\det (1- \p \wt x/\p x)} \, . 
\ee

In our case the action of $H$~on $AdS_{2} \times S^{2}$ decomposes into the 
separate actions of~$L_{0}$ and~$J_{0}$ on the~$AdS_{2}$ and~$S^{2}$ factors, respectively.
There are two fixed points -- at the center $\eta=0$ of the~$AdS_{2}$ 
factor (fixed by the rotation~$L_{0}$), and at the two poles on the~$S^{2}$ factor (fixed 
by the rotation~$J_{0}$). 
To apply the index theorem, we further need to know the explicit field content of~$X$ and~$\Psi$, and the 
charges they carry under~$H$. Once we know the eigenvalues of all the fields under~$H$, we can 
compute the trace in the numerator of~\eqref{ASindthm}. 
As we discussed in~\S\ref{offshellsusy}, the off-shell algebra that we use has the same structure as that used 
in~\cite{Pestun:2007rz,Hama:2012bg}, in that the field content and the gauge invariances are the same. 
This allows us to use the splitting of fields into~$X$,$\,\Psi$ as used by those authors. 
On the other hand, as was emphasized at the end of~\S\ref{offshellsusy}, the physical transformations 
on the right-hand side of the algebra as well as the background manifold are different, and we should 
use the algebra~\eqref{Qalg} that is relevant for our problem here. 

The action of the Lie derivative~$\CL_{v}$ on any field of the theory is composed of two parts: a local 
translation on the spacetime coordinates along the vector~$v^{\mu}$, and an action on the tensor 
indices of the field. At the fixed points of spacetime under~$H$, the former action vanishes by definition. 
Thus, in order to compute the action of~$H$, we only need to keep track of the latter action of the Lie derivative, 
as well as the action of the Lorentz rotation~$L_{ab}$. The vector~$v^{\mu}$ \eqref{vmu} translates us 
along the angles~$\tau$ and~$\phi$ in the metric~\eqref{metric2}
and is therefore essentially a rotation around the fixed points. 
The operator~$L_{ab}$~\eqref{ourlab} at the fixed points is also the same rotation (acting on the spin part of the fields). Therefore, we only need to compute the charges of the all fields under a rotation around the 
center of~$AdS_{2}$ combined with a rotation around~$S^{2}$.

The calculation is simplified by going to complex coordinates in which the~$AdS_{2} \times S^{2}$ metric is
\begin{equation}\label{metriccomplex}
    ds^2 \= \ell^{2} \biggl( \frac{4 dw d\bar w}{(1- w \bar w)^2} + \frac{4 dz d\bar z}{(1 + z \bar z)^2} \biggr) \, .
\end{equation}
Here~$\ell$ is the overall physical size of the~$AdS_{2} \times S^{2}$ metric, which is governed by the 
field-dependent physical metric~$e^{-\CK (X^{I})} g_{\mu\nu}$ that depends on the position in 
the~$AdS_{2}$ space. At the fixed points, i.e.~the center of~$AdS_{2}$, this size
is given by~$\ell^{2} = e^{-\CK (\phi^{I})}$ in the gauge $\sqrt{-g}=1$.\footnote{Here and in the following, 
we write~$\CK (\phi^{I})$ to mean~$\CK(\phi^{I}+ip^{I})$.} 
At the fixed points, we have~$w=0$, and~$z=0$ or~$1/z=0$. There, the action of the 
operator~$e^{-iHt}$ on the spacetime coordinates is~$(z,w) \to e^{- i t/\ell} (z,w)$.
Therefore, the determinant factor in the denominator of~\eqref{ASindthm} is~$(1-q)^{2} \, (1-q^{-1})^{2}$ 
with~$q = e^{- i t/\ell}$. 

Near the fixed points the space looks locally like~$\IR^{4}$ with an associated~$SO(4)=SU(2)_{+} \times SU(2)_{-}$ 
rotation symmetry. The planes labelled by  the two complex coordinates~$(z,w)$ rotate in the same direction under 
the~$SU(2)_{+}$, and in opposite directions under~$SU(2)_{-}$. Comparing the two forms of the metric~\eqref{metric} and~\eqref{metriccomplex} 
(noting the change in orientiation of the~$S^2$ part when going from one to another), and recalling that~$H = -i\partial_\tau + i\partial_\phi$, 
we identify the action of~$H$ with the Cartan generator of~$SU(2)_{+}$ at the North Pole, and with the Cartan of~$SU(2)_-$ at the South Pole according to: 
\be
H = 2\,J_+ \quad \text{(NP)}\, , \qquad H = 2\,J_- \quad \text{(SP)}\, .
\ee
We now need to compute the charges of all the fields under this generator.

\vspace{0.4cm}

\noindent \textbf{Vector multiplets:} 
In the vector multiplet sector, the fields are separated into~$X = \{ X^I-\bar{X}^I,A^I_\mu \}$ 
and~$\Psi=\{ \Xi^{I\;ij}\,, \,c^I,b^I \}$, and their images under~$\wh{Q}$. We discuss some more details of this 
splitting in Appendix \ref{symbol}. 
The fermions~$\Xi^{I\;ij}$ are defined as
\be \label{defxiij}
\Xi^{I\;ij} \,:= \,  2\, \bar{\xi}^{(i}_+\lambda^{I\;j)}_+ + 2 \, \bar{\xi}^{(i}_-\lambda^{I\;j)}_- \, . 
\ee
%

The scalars~$(X^I-\bar{X}^I)$, $c^{I}, b^{I}$ are in the~$(\bf{0},\bf{0})$ of~$SO(4) = SU(2)_+ \times SU(2)_-$, and therefore are 
uncharged under~$H$.
The vector rotates with spin one, and therefore is in the~$(\bf{2},\bf{2})$ of the $SO(4)$. 
There are two modes ($A_{z}$, $A_{w}$) with charges~$+1$ and two modes ($A_{\overline{z}}$, $A_{\overline{w}}$) with 
charges~$-1$ under~$H$.\footnote{Our convention is that a field~$\varphi$ of charge~$e$ transforms as~$\varphi \to e^{- i e H t} \, \varphi$.}
To compute the charges of the spinor bilinears, we notice that the spinor~$\xi_{+}$ vanishes at the north pole, and 
so the bilinear~$\Xi^{I\;ij}$ is in the~$(\bf{1},\bf{3})$ of the $SO(4)$. The spinor bilinears~$\Xi^{I\;ij}$ 
thus carry charge~$0$ under~$H$. Similarly, at the south pole, the spinor bilinears are in the~$(\bf{3},\bf{1})$, 
while~$H$ is the Cartan of the~$SU(2)_{-}$. All this is consistent with the explicit symbol computation 
in Appendix~\ref{symbol}, where the coupling of the bilinears with the self-dual and the anti-self-dual part of the 
field strength is computed. 



Putting all this together, we find that, at each of the poles, the contribution to the index is:
\be  \nonumber
\left[\frac{2q}{(1-q)^{2}}\right] \, . 
\ee
We see that there is a pole in this expression when~$q=1$. This pole arises due to the fact that our 
operator is not elliptic but transversally elliptic. At a hands-on level, the pole presents a problem in 
the interpretation of the index -- namely, how to compute the Fourier coefficients of this expression. 
Depending on whether we expand around~$q=0$ or~$q^{-1}=0$, we will obtain 
$2\sum_{n \ge 1} n \, q^{n}$ or $2\sum_{n \ge 1} n\,q^{-n}$, which clearly have different Fourier 
coefficients. This problem is resolved by giving a certain regularization defined by 
the behavior of the operator in the neighborhood of each fixed point~\cite{Pestun:2007rz}. Accordingly, we write: 
\be \label{vecind}
\text{ind}_{\text{vec}} (D_{10}) \= \left[\frac{2q}{(1-q)^{2}}\right]_{\text{NP}} + \left[\frac{2q}{(1-q)^{2}}\right]_{\text{SP}}\, . 
\ee
Here we have indicated the North Pole and South Pole contributions. 
As we shall see, the effect of the different regulators in our final results for the determinant will only be 
in an additive constant which we ignore in the functional determinant.

%

\vspace{0.4cm}
\noindent \textbf{Hyper multiplets:} We do a similar analysis for the hyper multiplets. The fields are separated 
into~$X=\{ A_i^{\;\alpha} \}$ and~$\Psi=\{ \Xi_i^{\;\alpha} \}$, with
\be
\Xi_i^{\;\alpha} := 2\epsilon_{ij}\left(\bar{\breve{\xi}}^{\,j}_+\zeta^\alpha_+ + \bar{\breve{\xi}}^{\,j}_-\zeta^\alpha_-\right) \, ,
\ee
again inspired by~\cite{Pestun:2007rz,Hama:2012bg}. Details of this field splitting can also be found in Appendix~\ref{symbol}.
The scalars~$A_i^{\;\alpha}$ do not transform under rotations. To compute the charges of the fermions, we note that now
it is the spinor~$\breve{\xi}_-$ that vanishes 
at the North Pole (as can be seen from the expression~\eqref{xicheck}), and therefore the spinor bilinear~$\Xi_i^{\;\alpha}$ 
is in the~$2\times(\bf{2},\bf{1})$ of~$SO(4)$, where the factor of~$2$ counts both~$\alpha$ components of a given hypermultiplet. 
Similarly at the South Pole,~$\breve{\xi}_+$ vanishes and therefore the bilinear 
is in the~$2\times(\bf{1},\bf{2})$ of~$SO(4)$. Putting everything together, we obtain the index for one hyper multiplet:
\be \label{hypind}
\text{ind}_{\text{hyp}} (D_{10}) \= \left[-\frac{2q}{(1-q)^{2}}\right]_{\text{NP}} + \left[-\frac{2q}{(1-q)^{2}}\right]_{\text{SP}} \, .
\ee

\vspace{0.4cm}

\subsection*{Zeta function regularization}

\noindent We now use the expressions~\eqref{vecind}, \eqref{hypind}, for the index of the vector and hyper multiplets, 
to compute their one-loop determinants. Given the infinite product~\eqref{detratio1}, we write a formal expression 
for the logarithm of the one-loop determinant as:
\be \label{sdetH}
\log \frac{\det_{\Psi} H}{\det_\text{X} H} \= -\sum_{n \ge 1} a(n)\,\log\l_{n} \, . 
\ee
In order to regularize this infinite sum, we use the method of zeta 
functions\footnote{The zeta function regularization has been used with great success 
to compute the perturbative one-loop corrections to the physical quantum gravity path integral 
(see~\cite{Hawking:1976ja} and follow-ups). 
Here we use the technique for the exact computation using localization methods.}. 
We first construct the zeta function:
\be
\zeta_{H} (s) \= \sum_{n \ge 1} \, a(n) \, \l_{n}^{-s} \, . 
\ee
This converges for suitably large values of~$\Re(s)$, and we then analytically continue it to the complex~$s$ plane. 
The superdeterminant~\eqref{sdetH} is then defined as:
\be \label{Hzeta}
\log  \frac{\det_{\Psi} H}{\det_\text{X} H}  \= \zeta_{H}'(s) \mid_{s=0} \, . 
\ee

One of the advantages of the zeta function method is that it easily yields the dependence of the determinant on 
the physical parameters of the problem. In our case, we have only one parameter in the background which is the 
overall size of the metric~$\ell^{2}=e^{-\CK(\phi^{I})}$. The dependence on~$\ell$ is easily calculated using the 
scaling properties of the zeta function~\cite{Hawking:1976ja}. 

We consider the contribution to the index at the north pole and at the south pole separately. 
At the north pole, we have an expression which is expanded around~$q=0$:
\be
\left[\frac{2q}{(1-q)^{2}}\right]_{\text{NP}} \= 2\sum_{n \ge 1}n\,q^n = \sum_{n \ge 1}2n\,e^{-it\tfrac{n}{\ell}} \, .
\ee
In the above language, this index has 
\be
a(n)\=2n \, , \quad \l_{n} \= \frac{n}{\ell} \, .
\ee
The zeta function for this piece of the determinant is 
\be
\zeta_{H}^\text{NP} (s) \=\sum_{n \ge 1} 2n \, \bigl(\frac{n}{\ell}\bigr)^{-s} \=  2 \, \ell^{s} \, \zeta_{R}(s-1) \, , 
\ee
where we have introduced the Riemann zeta function 
\be
\zeta_{R}(s) \= \sum_{n \ge 1} \, \frac{1}{n^{s}} \, .  
\ee
At the south pole, where we expand in powers of~$q^{-1}$, we get a similar expression but the zeta 
function~$\zeta_{H}^\text{SP} (s) $ there differs from the north pole answer by a factor of~$(-1)^s$. 
We thus need to deal with expressions of the type~$\log(-n)$, for which we use the positive branch of the logarithm.  

Putting together the north pole and the south pole contributions, we obtain 
\bea \label{zetaHans}
\zeta_{H}'(s) \mid_{s=0} & = & 4 \zeta'_{R}(-1) + 2\pi i \zeta_R(-1) + 4 \, \zeta_{R}(-1) \, \log \ell \nonumber \\
 & = & 4 \zeta'_{R}(-1) + 2\pi i \zeta_R(-1) + \frac{1}{6} \, \CK (\phi^{I}) \, . 
\eea
Since we are not keeping track of purely numerical overall constants, we drop the finite 
numbers~$4 \zeta'_{R}(-1)$ and~$2\pi i\zeta_R(-1)$ in further expressions. 
Putting together Equations~\eqref{Z1loop}, \eqref{detratio}, \eqref{Hzeta}, and \eqref{zetaHans}, we obtain:
\be \label{1loopdetvec}
Z_\text{1-loop}^\text{vec} (\phi^{I}) \= \exp\bigl(\CK(\phi^{I})/12 \bigr) \, , 
\ee
with~$\CK(\phi^{I})$ the generalized K\"ahler potential defined in Equation~\eqref{EminK}.

For the hyper multiplets, we use the same technique, and we find that the index is equal and opposite to that
of the vector multiplet -- as can be seen directly from the expressions~\eqref{vecind},~\eqref{hypind}. 
Our final result is:
\be \label{1loopdet}
Z_\text{1-loop}^\text{vec} (\phi^{I}) \= \left(Z_\text{1-loop}^\text{hyp} (\phi^{I})\right)^{-1} \= \exp\bigl(\CK(\phi^{I})/12 \bigr) \, . 
\ee

\vspace{0.2cm}

Although we have only worked out the details of the vector and hyper multiplets, it is clear that 
the above calculation will also go through essentially unchanged once we have fixed the off-shell complex of any multiplet. 
Since there is only one scale set by~$e^{-\CK}$ in the localization background, 
the functional determinant will have the symplectically invariant form~$e^{-a_{0} \CK(\phi^{I})}$. 
The number~$a_{0}$ receives contributions from each multiplet of the~$\CN=2$ supergravity theory:
\be \label{defa0}
a_{0} \= a_{0}^\text{grav} \+ n_{3/2} \, a_{0}^{3/2} \+ (\nv + 1) \, a_{0}^\text{vec} \+  \nh \, a_{0}^\text{hyp} \, , 
\ee
where~$n_{3/2}$, $(\nv+1)$, $\nh$ are the number of gravitini, vectors and hypers in the off-shell theory, respectively. 
From our results in this section,~$a_{0}^\text{vec}=-a_{0}^\text{hyp}=-1/12$.
As mentioned in the introduction, we are currently investigating the details of the off-shell computation of 
the graviton and gravitini determinants~\cite{deWitMurReys}. 
We will see in the following section how we can use existing on-shell computations to check our formula~\eqref{1loopdet}
for the vectors and hypers, as well as to deduce the coefficients~$a_{0}$ for the other multiplets.

\section{Relation to previous results for the black hole entropy \label{Expansions}}

The leading logarithmic corrections to the classical black hole entropy have been obtained 
in~\cite{Banerjee:2010qc, Banerjee:2011jp, Sen:2011ba} by explicitly evaluating the one-loop determinant 
of the kinetic terms of all the quadratic fluctuations of the theory around the classical attractor background~\eqref{metric}. 
This is a very intricate computation that needs a diagonalization of the kinetic terms of all the fields of the theory, and 
it depends on the fact that the values of the metric, fluxes and scalars in the attractor solution are related by 
supersymmetry\footnote{The recent interesting work of~\cite{Keeler:2014bra, Larsen:2014bqa} also 
uses on-shell techniques, but relies only on the chiral spectrum of the supersymmetry algebra. }. 
In contrast, the localization method involves the one-loop determinant of the deformation operator~$Q\CV$, which does not 
depend on the equations of motion and the associated kinetic mixings. 
At a practical level, the on-shell computation of~\cite{Banerjee:2010qc, Banerjee:2011jp, Sen:2011ba} 
proceeds by solving for the spectrum of eigenvalues of the various Laplacians of the theory,
and observing that there is a huge cancellation among them. The index theorem, on the other hand, 
reduces this problem to a very simple computation at the fixed points of a certain~$U(1)$ action. 

The results of the on-shell and off-shell methods agree in the large-charge limit, as expected. 
In fact a little more can be said about the 
interplay of the techniques used in these two approaches. In this section we present our understanding of 
this relationship. Using these relations, we also explain the cancellations regarding $\frac18$-BPS black holes in~$\CN=8$ string theories
that were observed in~\cite{Dabholkar:2011ec}.

\subsection*{Relation to large-charge on-shell computations}

We consider a limit in which all the charges~$(q_{I}, p^{I})$ scale uniformly by 
a large parameter~$\Lambda$, i.e.~$(q_{I}, p^{I}) \to \Lambda (q_{I}, p^{I})$. 
In the leading $\Lambda \to \infty$ limit, one can evaluate the quantum entropy integral~\eqref{integral} 
using the saddle point method.
If we ignore the determinant factor~$Z_\text{det}$, the saddle point equations are simply
the extremization equations of the exponent of~\eqref{integral}. As we discussed in~\S\ref{formalism}, 
these extremization equations are precisely the attractor equations~\eqref{attraceq}, and 
the saddle point values~$\phi^{I}_{*}= \Re X^{I}_{*}$, the attractor value of the scalar fields. 

From the attractor equations~\eqref{attraceq}, we see that the attractor values~$\phi^{I}_{*} \sim \Lambda$ 
for large~$\Lambda$, and the attractor entropy~\eqref{Sattr} scales as~$\Lambda^{2}$. 
From Equation~\eqref{EminK}, we see that the determinants~\eqref{1loopdet} scale 
as~$\Lambda^{-2a_{0}}$ and therefore they will contribute to the entropy as~$\log{\Lambda}$,  
so that it is indeed justified to ignore them to leading order. The resulting classical entropy is: 
\be \label{SclassAgain}
S^\text{class}_\text{BH}\,  \= \pi \bigl(-q_{I} \,e^{I}_{*} - 4 \, \Im F^{(0)} ((e^{I}_{*}+ i p^{I})/2) \bigr) 
\, \approx \, \frac{A_{H}}{4} \, , 
\ee
where $F^{(0)}$ denotes the prepotential without any $\wh{A}$-dependence, corresponding to the two-derivative 
effective action, which is consistent with the large-charge approximation. This entropy agrees with the attractor 
mechanism~\eqref{Legendre}, \eqref{Sclass}.

The first corrections to the leading large-charge entropy are given by the first corrections to the 
saddle point value~\eqref{SclassAgain}, of~\eqref{integral}.
In the large-charge limit, we know that~$A_{H} \sim e^{-\CK} \sim \Lambda^{2}$. From Equation~\eqref{1loopdet}
we deduce that
\be \label{leadlog}
\SBH \= \frac{A_{H}}{4} + a_{0} \log A_{H} \+ \cdots \, ,
\ee
where the number~$a_{0}$ is precisely the coefficient defined in~\eqref{defa0}. In~\S\ref{DetCalc}, 
we saw that
\be \label{a0vechyp}
a_{0}^\text{vec} \= - a_{0}^\text{hyp} \= -\frac{1}{12}  \, , 
\ee
which indeed agrees with the corresponding on-shell computations of the log corrections to 
the black hole entropy \cite{Sen:2011ba}, performed using the heat-kernel method. In the rest of this section,
we make some comments on the relation between our exact index calculation of \S\ref{DetCalc}, 
heat-kernels, and the large-charge expansion.

The heat-kernel method (see e.g.~\cite{Vassilevich:2003xt}) to compute the functional determinant of an operator~$D$ 
uses its representation as an integral over the proper time~$t$:
\be \label{heatker}
\half \log \det(D) \= - \half \int_{0}^{\infty} \frac{dt}{t} \, K(t,D) \, , \qquad 
 K(t,D) \= \Tr \, e^{-Dt} \, . 
\ee 
The integral on the right-hand side of~\eqref{heatker} is not always well-defined. 
The divergences as~$t \to 0$ arises from the UV divergences, for which we assume a UV cutoff~$\ve$. 
The divergences as~$t \to \infty$ appear because of zero or negative eigenvalues 
of~$D$. In our problem, the relevant operator~$H$ does not have any negative eigenvalues, nor does it 
have zero modes since, due to the boundary conditions we impose, the ghost-for-ghost fields are absent. 

The coefficient of the logarithmic term in a large-charge expansion of quantum black hole entropy~\eqref{qef} 
is determined by the constant coefficient in the~$t \to 0$ expansion of the integrated heat-kernel~\cite{Sen:2011ba}.  
Our calculations of \S\ref{DetCalc} can be written as (here~$q=e^{-t}$):
\be \label{hypdet}
\half \log \text{det}_\text{vec}(H) \= - \half \log \text{det}_\text{hyp}(H) \= \int_{\ve}^{\infty}  \frac{dt}{t} \frac{2q}{(1-q)^{2}}  \, . 
\ee
If we are only interested in the logarithmic term in the large charge expansion, we can also directly using the~$t \to 0$ 
expansion of the heat kernel in the above integrals: 
\be
\frac{2q}{(1-q)^{2}} \= \frac{2}{t^2}-\frac{1}{6} + \frac{t^{2}}{120}+O(t)^4 \, ,
\ee
from which we recover the result~\eqref{a0vechyp} for the coefficient~$a_{0}$.

We defined the number~$a_{0}$ as appearing in the off-shell one-loop determinant in~\S\ref{DetCalc}, and 
we saw above that the same number is the coefficient of the logarithmic correction to the large-charge expansion
of black hole entropy. 
We can actually use this consistency between on-shell and off-shell methods to deduce the value of~$a_{0}$
for the graviton and gravitini multiplets. The results of~\cite{Sen:2011ba} demand that~$a^{3/2}_{0}=-\frac{11}{12}$, 
and~$a^\text{grav}_{0}=2$ in the gauge $\sqrt{-g}=1$.

\subsection*{Miraculous cancellation in truncations of extended supergravities}

In~\cite{Dabholkar:2011ec}, the $\frac18$-BPS black hole in $\CN=8$ theory was considered from both the 
macroscopic and microscopic point of view. 
The physical low energy macroscopic field content is that of an~$\CN=8$ graviton multiplet which, in the 
$\CN=2$ language that we are considering, consists of one $\CN=2$ graviton multiplet, $n_{3/2}=6$ gravitini 
multiplets, $\nv=15$ vector multiplets, and $\nh=10$ hyper multiplets. The macroscopic entropy was computed using 
localization in~\cite{Dabholkar:2011ec} in the \emph{truncated} theory that was first considered in~\cite{Shih:2005he}, 
where the physical spectrum consists only of the $\CN=2$ graviton multiplet coupled to $\nv^\text{trun} =7$ vector multiplets. 

In this truncated theory, only the measure for the zero-modes of~$Q$ was taken into account 
in~\cite{Dabholkar:2011ec}, and it was computed to 
be~$Z_{0} =  e^{(\nv^\text{trun} +1)\CK/2} \times O(\Lambda^{0})$.
Assuming further that the non-zero mode determinant~$Z'_\text{det}=1$  the formula derived was: 
\be
W^\text{pert}(\Delta) =  \sqrt{2}  \, \pi \, \frac{1}{\Delta^{7/4}} \, I_{7/2}(\pi \sqrt{\Delta}) \, , 
\ee
where~$\Delta$ is the unique quartic U-duality invariant of the charges of the~$\CN=8$ theory. 
This formula was seen to agree on the nose with the microscopic formula for the black hole 
degeneracy~\cite{Dabholkar:2011ec}. 

We now have a better understanding of this agreement. 
Let us split the contribution of one vector multiplet into two parts 
as~$a_{0}^\text{vec}=-\frac{1}{12} = -\half+\frac{5}{12}$, where the~$-\frac{1}{2}$ is the contribution considered 
in~\cite{Dabholkar:2011ec}, and~$\frac{5}{12}$ is the rest. Then, using the values of~$a_{0}$ for the 
various multiplets written in the previous subsection, the contribution to~$a_{0}$ ignored 
in~\cite{Dabholkar:2011ec} is 
\be\nonumber
\frac{5}{12}  (\nv^\text{trun}+1) -\frac{1}{12} (\nv - \nv^\text{trun}) + \frac{1}{12} \nh -\frac{11}{12} n_{3/2} + 2 \, . 
\ee
For the field content of the~$\CN=8$ theory and the~$\CN=2$ truncation as given above, this indeed adds 
up to zero, thus explaining the miraculous cancellation in the full string theory seen in~\cite{Dabholkar:2011ec}. 
This cancellation can already be seen at the leading log level from the results of~\cite{Sen:2011ba}.
It is now clear from the comments in this section that this cancellation holds to all orders in perturbation theory.

We can also consider~$\CN=4$ string theories, where the physical low energy macroscopic field content is 
an~$\CN=4$ graviton multiplet coupled to~$N_\text{v}$ $\CN=4$ vector multiplets. In terms of~$\CN=2$ multiplets, 
we have one graviton multiplet, $n_{3/2}=2$ gravitini multiplets, $\nv=N_\text{v}+1$ vector multiplets, 
and $\nh=N_\text{v}$ hyper multiplets. 
The total logarithmic correction according to~\eqref{defa0} is given by~$a_{0} = 2-\frac{11}{12}\times 2 - \frac{1}{12} \times 2 =0$, 
as consistent with the on-shell computations in the limit when all the charges are scaled to be equally large. 
We can also consider a truncation in which we have an agreement for the 
leading Bessel function in the Cardy limit~\cite{DGMR}.

\section{Exact formulas for $\CN=2$ quantum black hole entropy and the relation to topological strings \label{ExFor}}

The true power of the localization method clearly lies in the fact that one can go beyond the perturbative large charge 
approximation to get an exact result for black hole entropy. In this section we propose such an exact 
entropy formula for BPS black holes in~$\CN=2$ supergravity coupled to~$\nv$ vector multiplets and~$\nh$ 
hyper multiplets. We then make some comments relating our formula to the microscopic formula 
of~\cite{Denef:2007vg}, as well as on some relations with topological string theory.

In the previous sections, we have seen that one-loop determinant of the fluctuations around the localization manifold 
takes the symplectically invariant form\footnote{In this section we assume~$a_{0}^\text{grav} = 2$ (as argued for above) 
in the gauge $\sqrt{-g}=1$ which we use throughout this paper. It is important to derive this result from a proper analysis 
of the fluctuating Weyl multiplet and the corresponding gauge-fixing. This is under investigation \cite{deWitMurReys}.}:
\be \label{1loopfinal}
Z_\text{1-loop} \= \exp\Bigl(-\CK(\phi^{I}) \bigl(2 -\frac{\chi}{24} \bigr) \Bigr) \, , \qquad \chi \= 2(\nv+1-\nh) \, .
\ee
Recall from the discussion below Equation~\eqref{integral} that the full determinant factor in the exact formula 
has two contributing pieces -- the one-loop fluctuations~$Z_\text{1-loop}$, and the measure from the curvature of field space 
itself~$Z^\text{ind}_{\text{det}}$. Combining these elements, we obtain:
\be \label{Wpert}
W^\text{pert} (q, p) = \int_{\mathcal{M}_{Q}}  \, \prod_{I=0}^{\nv} d\phi^{I} \, 
e^{- \pi  \, q_I  \, \phi^I  + 4 \pi \, {\rm Im}{F ((\phi^I+ip^I)/2)}}
e^{-\CK(\phi^{I}) (2 - \chi/24) } \, 
Z^\text{ind}_{\text{det}} \, . 
\ee

To move on, we need to discuss the details of the prepotential function~$F(X^{I},\wh{A})$,
which is a holomorphic homogeneous function of weight 2 in its variables under the scalings~$X^{I} \to \lambda X^{I}$, 
$\wh{A} \to \lambda^{2} X^{I}$. 
Generically, we have an expansion of the form:
\be \label{FAexp}
F(X^{I},\wh{A}) \= \sum_{g=0}^{\infty} \, F^{(g)}(X^{I}) \, \wh{A}^{g} \, 
\ee
that enters the Wilsonian effective action 
of the on-shell supergravity. The function~$F^{(0)}(X^{I})$ controls the two-derivative interactions, 
and the coefficients~$F^{(g)}$, $g \ge 1$, describe higher derivative couplings of the form~$C^{2} \, T^{2g-2}$ and 
terms related by supersymmetry, where~$C$ is related to the Weyl tensor, and~$T$ is related to the 
graviphoton field strength.

At the two-derivative level, the prepotential has the form
\be \label{classprep}
F^{(0)}(X^{I}) \= -\half \sum_{i,j,k=1}^{\nv} C_{ijk} \, \frac{X^{i} X^{j} X^{k}}{X^{0}} \, , 
\ee
for a choice of symmetric~$C_{ijk}$. At this level, we can think of the measure of the scalars in a geometric manner, and 
compute it from the knowledge of the two-derivative kinetic term of the scalar sigma model. To be more thorough, 
we should take into account all the fields in the theory -- this can be done by using 
duality invariance as a criterion for the measure as in~\cite{LopesCardoso:2006bg}. 
Both these approaches give rise to the measure:
\be \label{indmsr0}
Z^\text{ind}_\text{det} \= (\det \text{Im}(F^{(0)}_{IJ}))^{\frac12} \; . 
\ee

For a prepotential of the form~\eqref{classprep}, and for\footnote{In the type IIA setting, 
this means absence of D6-branes in the charge configuration making up the black hole.}~$p^{0}=0, q_{0}\neq 0$, 
we can compute the various expressions entering the exact formula~\eqref{Wpert}.  We have:
\be \label{EminKatt}
e^{-\CK^{(0)}} \= \frac{C_{ijk} \, p^{i} \, p^{j} \, p^{k}}{\phi^{0}} \, ,
\ee
and~$\det \text{Im}(F^{(0)}_{IJ}) = A/(\phi^{0})^{(\nv+3)/2}$ where~$A$ does not depend on~$\phi^{I}$ (but does depend 
on~$C_{ijk}$ and~$p^{i}$). 
However, using these expressions in our integral expression~\eqref{Wpert}
leads to a formula which does not match the corresponding microscopic BPS state counting formulas beyond 
the leading logarithmic correction (see e.g.~\cite{Dabholkar:2005dt, Pioline:2006ni, Denef:2007vg}). 

From our point of view, this discrepancy arises from our lack of complete understanding of the induced measure term. 
The current best understanding of the measure in the supergravity field space comes from the 
work of~\cite{LopesCardoso:2006bg, Cardoso:2008fr}, whose main guiding principle is duality invariance. 
These authors have argued that imposing duality invariance leads to a non-holomorphic modification to 
the induced measure. At the two-derivative level, including these corrections, one has:
\be \label{indmsr}
Z^\text{ind}_\text{det} \= \bigl( \phi_{0}^{-2} \, \exp(-\CK^{(0)}(\phi^{I})) \bigr)^{\frac{\chi}{24}-1} \; , 
\ee
We note that the precise context in which these modifications have been derived 
is different from the one considered here. Notwithstanding this difference, if we 
combine the expression~\eqref{indmsr} and the one-loop factor~\eqref{1loopfinal} 
in our exact formula~\eqref{Wpert}, we obtain:
\be \label{integral2}
W^\text{pert} (q, p) = \int_{\mathcal{M}_{Q}}  \, \prod_{I=0}^{\nv} d\phi^{I} \, \exp\Big(- \pi  \, q_I  \, \phi^I 
 + 4 \pi \, \Im{F^{(0)} \big((\phi^I+ip^I)/2 \big)} \Big) (\phi^{0})^{2-\frac{\chi}{12}} \, e^{-\CK^{(0)}(\phi)} \, .
\ee
The black hole entropy formula conjectured in the paper~\cite{Denef:2007vg} based on  
consistency with the Rademacher expansion of the \emph{microscopic} black hole degeneracies in string theory
has exactly the same form as~\eqref{integral2}, with the two-derivative expressions~$F^{(0)}$, $\CK^{(0)}$ replaced
by the all-order expressions~$F$, $\CK$, respectively. 

To go beyond the two-derivative level in our formalism, we need a formula for the induced measure 
at all orders. The work of~\cite{LopesCardoso:2006bg, Cardoso:2008fr} provides a formalism to take into 
account all the holomorphic corrections to the supergravity measure. More work, however, needs to be 
done to fully understand the non-holomorphic effects in the induced measure as defined in our treatment.  
It is possible that the \emph{a priori} induced measure in the original supergravity 
path integral suffers from a holomorphic anomaly. 
Similar ideas have been proposed in~\cite{Verlinde:2004ck} in the context of the topological string theory. 
A computation of this a priori measure from first principles would complete
the derivation of the exact quantum black hole entropy in the gravitational theory.

\subsection*{Comments on relations to topological string theory}

Consider type IIA string theory compactified on a Calabi-Yau 3-fold~$CY_{3}$. 
The A-model topological string partition function on~$CY_{3}$ has the expression:
\be \label{Ftop}
F_\text{top} \= -i \frac{(2\pi)^{3}}{6\lambda^{2}} \, C_{ijk} \, t^{i} \, t^{j} \, t^{k} - \frac{i \pi}{12} \, c_{2i} \, t^{i} 
+ F_{GW}(\lambda, t^{i}) \, , 
\ee
where~$\lambda$ is the topological string coupling, $t^{i}$ are the moduli fields (the complexified K\"ahler 
structure in the type IIA theory), $c_{2i}$ are the second Chern classes of the 4-cycles of the~$CY_{3}$, 
and~$F_{GW}$ is the generating function of the Gromov-Witten (GW) invariants of the~$CY_{3}$ that admits 
an expansion in powers of~$\lambda$. 
By comparing~\eqref{Ftop} to the corresponding Wilsonian expression~\eqref{FAexp} in the supergravity,
we obtain\footnote{There are important subtleties associated with the above identification, having to do with 
the action of duality (symplectic transformations) on the geometry of the Calabi-Yau surface and in 
supergravity~\cite{Cardoso:2008fr, Cardoso:2014kwa}. We do not add anything to this discussion.}: 
\be
F_{\text{top}}\=\frac{i \pi}{2} F, \qquad   t^{i}\= \frac{X^{i}}{X^{0}} \, , 
\qquad \lambda^{2} \= \frac{\pi^{2}}{8} \frac{\wh A}{(X^{0})^{2}} \, . 
\ee

The value of the topological string coupling constant on the supergravity localization manifold 
analyzed in this paper is $|\lambda| = 2 \pi \sqrt{2}/\phi^{0}$ -- which is small for large values of the charges. 
The microscopic analysis of~\cite{Dabholkar:2005dt, Pioline:2006ni, Denef:2007vg} is based on large~$\lambda$. 
Using the relation of the GW invariants to the Gopakumar-Vafa invariants related to counting M2-branes in M-theory,
then making a precise prediction for the degenerate instanton contribution at large topological string coupling,
and a subsequent analytic continuation, the authors of~\cite{Dabholkar:2005dt, Pioline:2006ni, Denef:2007vg} 
claimed that the 
the topological string partition function at weak coupling must have an additional logarithmic term:
\be \label{Ftop2}
\wt F_\text{top} \= -i \frac{(2\pi)^{3}}{6\lambda^{2}} \, C_{ABC} \, t^{A} \, t^{B} \, t^{C} -  \frac{\chi}{24} \log\lambda 
- \frac{i \pi}{12} \, c_{2A} \, t^{A} + F_{GW}(\lambda, t^{A})  \, .
\ee
where~$\chi$ is the Euler characteristic of the Calabi-Yau three-fold.
The puzzle then is to interpret the logarithmic term in supergravity. 
Being a non-local contribution, it cannot arise at any order in perturbation theory in~$\wh{A}$.

From our point of view, the logarithmic contribution in~$\lambda$ (or equivalently in~$\wh{A}$) 
appears as a quantum effect. If we interpret the formula~\eqref{integral2} as an OSV type formula, 
then the imaginary part of the prepotential contains precisely the additional non-local logarithmic piece 
with coefficient~$\chi/24$ that is predicted by the analytic continuation of the microscopic theory. 
(We recall that in a string compactification on a~$CY_{3}$, the number~$\chi=2(\nv+1-\nh)$ is the Euler 
characteristic of the~$CY_{3}$.)
Our~$AdS_{2}$ functional integral incorporates the integration 
over massless modes, and although the Wilsonian action of supergravity does not contain the logarithmic term, 
the effective 1PI action appearing in the exponent of Equation~\eqref{integral2}  
does.\footnote{A deeper explanation of this phenomenon appears in~\cite{Dedushenko:2014nya}.}$\,$\footnote{There are similar $\log g_s$ terms in the couplings of the low energy effective action 
of string theory in flat space, e.g.~\cite{Kiritsis:1997em}, which can be explained by mixing between the local 
and non-local part of the 1PI action when rescaling from string frame to Einstein frame~\cite{Green:2010sp}.}
We mention that most of this interpretation can be reconstructed by combining the duality arguments 
of~\cite{Cardoso:2008fr, Cardoso:2014kwa} with the computation of the leading logarithmic effects 
of~\cite{Sen:2011ba}. The one point we add to this discussion is the direct calculation of the one-loop 
effects proportional to~$e^{-a_{0}\CK}$.

\vspace{0.2cm}

Finally, we note that, in addition to being at different values of coupling constants, the values of the moduli 
in our analysis and that of~\cite{Denef:2007vg} are also different. The authors of~\cite{Denef:2007vg} work 
with moduli~$t_{\infty}$ in asymptotically flat space, while we choose attractor values of moduli to 
define the black hole degeneracy since we are only interested in the single-center black holes. 
Our results could be interpreted to mean that the relevant index does not suffer any wall-crossing 
on moving from one regime to the other. 

These results may also point to a new ``black hole index'' that is simply constant over all of moduli space. 
Indeed, an argument was made in~\cite{Sen:2009vz, Dabholkar:2010rm} that, when a black hole preserves at least four 
supercharges and consequently at least an~$SU(2)_{R}$ symmetry at its horizon, its quantum entropy is equal 
to a supersymmetric index. Defining this index in the microscopic theory is not an easy problem, but one can 
do that in~$\CN=4$ string theories. In that case the black hole index is given by the coefficient of a mock modular form, 
defined using the attractor value of moduli, and it is constant all over of moduli space~\cite{Dabholkar:2012nd}. 
A similar phenomenon in~$\CN=2$ string theories would point towards a larger symmetry underlying the BPS states 
of~$\CN=2$ theories as proposed in~\cite{Harvey:1995fq, Harvey:1996gc}.

\section*{Acknowledgements}
We would like to thank Daniel Butter, Stefano Cremonesi, Bernard de Wit, Atish Dabholkar, Jo\~ao Gomes, 
Mahir Hazdic, Gianluca Inverso, Sungjay Lee, Hirosi Ooguri, Boris Pioline, Ashoke Sen,
and especially Rajesh Gupta and Imtak Jeon for useful and informative conversations. 
This work is supported by the ERC Advanced Grant no.~246974, {\it ``Supersymmetry: a window to non-perturbative physics''} 
and the EPSRC First Grant UK EP/M018903/1.


\appendix

\section{Conventions  \label{euclideanspinors} } 

The summation convention for~$SU(2)$ indices is NW-SE and (anti)symmetrization of indices is done with weight one. The antisymmetric tensor of~$SU(2)$ is such that
\be
\epsilon^{ij}\epsilon_{jk} = -\delta^i_k \quad \text{and} \quad \epsilon^{ij}\epsilon_{ij} = 2\, .
\ee

We take the following hermitian tangent space Dirac matrices in Euclidean signature:
\be \label{cliffordrep}
\gamma_1 = \sigma_1 \otimes 1\, , \quad \gamma_2 = \sigma_2 \otimes 1\, , \quad 
\gamma_3 = \sigma_3 \otimes \sigma_1\, , \quad \gamma_4 = \sigma_3 \otimes \sigma_2\, ,
\ee
where~$\sigma_i$,~$i=1,2,3$ are the Pauli matrices. We also define the usual combination~$\gamma^{ab} = \tfrac{1}{2}[\gamma^a,\gamma^b]$ and similarly for higher-rank~$\gamma$ matrices. In addition,~$\gamma_5=-\gamma_1\gamma_2\gamma_3\gamma_4$. These matrices obey the following useful identities in four dimensions:
\bea
\gamma_{abc} &=& \epsilon_{abcd}\gamma^d\gamma_5 \, , \qquad \gamma_{abcd} = -\epsilon_{abcd}\gamma_5\, , \cr
\gamma^{ab}\gamma_b &=& 3\gamma^a \, , \qquad \quad \;\;\, \gamma^{abc}\gamma_c = 2\gamma^{ab}\, , \\
\gamma_b\gamma^a\gamma^b &=& -2\gamma^a \, , \quad \;\;\;\;\; \gamma_c\gamma^{ab}\gamma^c = 0\, , \cr
\gamma_a\gamma_{bc} &=& \delta_{ab}\gamma_c - \delta_{ac}\gamma_b + \epsilon_{abcd}\gamma^d\gamma_5\, . \nonumber
\eea

We define the Majorana conjugate of a spinor~$\lambda$ as
\be
\bar{\lambda} \equiv \lambda^T \mathcal{C}\, ,
\ee
where the charge conjugation matrix is given by~$\mathcal{C} \equiv \sigma_2 \otimes \sigma_1$ in the Clifford algebra representation~\eqref{cliffordrep}. For a Majorana spinor carrying an extra~$SU(2)$ index (e.g. because of R-symmetry), the symplectic Majorana reality condition reads
\be
\left(\lambda^i\right)^* = \epsilon_{ij}\,\mathcal{C}^T\lambda^j \, .
\ee
This condition can be imposed consistently in 4-dimensional Euclidean space, along with a chirality (Weyl) projection. 

For two spinors~$\chi$ and~$\lambda$, we have the so-called ``Majorana flip relation'':
\be
\bar{\chi}\,\gamma_{(r)}\lambda = \pm\,t_r\,\bar{\lambda}\,\gamma_{(r)}\chi\, ,
\ee
where~$\gamma_{(r)}$ is a Dirac matrix of rank~$r$, and the plus sign holds when \emph{both}~$\chi$ and~$\lambda$ are anti-commuting (Grassmann-odd). In 4-dimensional Euclidean space, we take
\be
t_0 = 1 \, , \quad t_1 = -1 \, , \quad t_2 = -1 \, , \quad t_3 = 1 \quad \text{and} \quad t_{r+4}=t_r \, .
\ee

We denote the (anti)self-dual part of an~$SU(2)$ antisymmetric tensor as
\be
T^-_{ab} = T_{ab}^{ij}\,\epsilon_{ij}\, , \qquad T^+_{ab} =T_{ab\,ij}\,\epsilon^{ij}\, .
\ee
A useful property of spinors and antisymmetric tensors is that when~$T_{ab}\gamma^{ab}$ acts on a spinor of (positive) negative 
chirality, it is projected onto its (anti)self-dual part:
\be
T_{ab}\gamma^{ab}\xi^i_+ = T^-_{ab}\gamma^{ab}\xi^i_+ \quad \text{and} \quad T_{ab}\gamma^{ab}\xi^i_- = T^+_{ab}\gamma^{ab}\xi^i_-\, .
\ee

\subsection*{Analytic continuation to Euclidean space}


The $\CN=2$ superconformal algebra that we use~\cite{deWit:1979ug, deWit:1980tn} holds for theories in Minkowski signature. 
We need to adapt it to our problem of computing a Euclidean path integral, for which we follow the approach 
of~\cite{Pestun:2007rz}. The idea is to use the original Minkowski algebra to perform algebraic computations such as computing 
the field variations, the action, and the symbol in Appendix~\ref{symbol}, and then perform an analytic continuation to Euclidean space, which 
we describe below, at the very end of the algebraic computations. In this paper, we have indicated this procedure by the inner product~$\left(.\, , \,.\right)$, 
and some explicit formulas are given in the symbol computation. 
This procedure could be streamlined by directly developing a Euclidean algebra 
from the beginning, this problem is currently being addressed. 

The Euclidean continuation is performed via the usual~$t\rightarrow i\tau$ in the metric and field configuration, and we regard Minkowski complex conjugates as independent fields in Euclidean space, i.e.~$X^I$ and~$\bar{X}^I$ and are two independent scalars,~$\Omega_i$ is independent of~$\Omega^i$, and so on. Moreover, in 4-dimensional spacetime with Minkowski signature, fundamental spinors are Weyl or Majorana spinors, whereas in Euclidean signature they are symplectic Majorana-Weyl~\cite{VanProeyen:1999ni}. To accommodate for this change of reality property when continuing~$\CN=2$ superconformal gravity to Euclidean signature, we follow the method of~\cite{Dabholkar:2010uh} and introduce new spinors~$\xi^i_\pm$ to parametrize the supersymmetry transformations, where~$i=1,2$ is an~$SU(2)$ index and~$\pm$ denotes the chirality of the spinor. S-supersymmetry is also parametrized with Euclidean spinors~$\eta^i_\pm$. We replace the Minkowski spinor parameters entering the algebra according to:
\be
\epsilon_i \rightarrow \epsilon_{ij}\xi^j_- \; ; \; \epsilon^i \rightarrow \xi^i_+ \, ,\qquad \text{and} \qquad \eta_i \rightarrow \epsilon_{ij}\eta^j_+ \; ; \; \eta^i \rightarrow \eta^i_- \, .
\ee
By definition, these spinors obey the symplectic Majorana condition
\be \label{AppSM}
(\xi^i_\pm)^* = -\epsilon_{ij}(\sigma_2\otimes\sigma_1)\xi^j_\pm \, ,
\ee
and similarly for~$\eta^i_\pm$. In this paper, we will take the spinors parametrizing the supersymmetry transformations to be Grassmann-even (commuting) spinors. This can be achieved by extracting a Grassmann number on both sides of the transformation rules. We may then consider successive supersymmetry transformations with equal parameters directly. This is useful since, as explained in the introduction of this paper, we will be interested in writing the action of the supercharge~$Q$ generated by a specific parameter, chosen so that~$Q$ squares to~$(L_0-J_0)$ which is the relevant supersymmetry for localization.

\vspace{.2cm}

\ndt \emph{Vector multiplets:} In a similar fashion, we introduce new Euclidean spinors~$\lambda^i_\pm$ to analytically continue the spinors of the vector multiplets (here we do not extract a Grassmann number from the spinors so the~$\lambda^i_\pm$ are still anti-commuting)
\be
\Omega_i^I \rightarrow \epsilon_{ij}\lambda^{I\;j}_+ \qquad \text{and} \qquad \Omega^{I\;i} \rightarrow -\lambda^{I\;i}_- \, .
\ee

\ndt \emph{Hyper multiplets:} We introduce the following spinors~$\zeta^\alpha_\pm$ to continue the hyperini (which are also still anti-commuting)
\be
\zeta^\alpha \rightarrow \zeta^\alpha_- \qquad \text{and} \qquad \zeta_\alpha \rightarrow \rho_{\alpha\beta}\,\zeta^\beta_+ \, ,
\ee
where~$\rho_{\alpha\beta}$ is a skew-symmetric constant matrix~\cite{deWit:1984px} satisfying
\be
\rho_{\alpha\beta}\rho^{\beta\gamma} = -\delta^\gamma_\alpha \quad \text{with} \quad \rho^{\alpha\beta} \equiv \left(\rho_{\alpha\beta}\right)^* \, .
\ee

\section{Killing spinors of~$AdS_2\times S^2$ \label{AppKilling}}

In this appendix, we review the solutions to the BPS equations of~$\CN=2$ supergravity:
\be \label{AppCKS1}
2D_\mu\xi^i_\pm \pm \tfrac{1}{16}T_{ab}^\mp\gamma^{ab}\gamma_\mu\xi^i_\mp - \gamma_\mu\eta^i_\mp = 0 \, ,
\ee
\be \label{AppCKS2}
\gamma^\mu D_\mu T^\mp_{ab} \gamma^{ab}\xi^i_\mp \pm 24 D\xi^i_\pm - T^\mp_{ab}\gamma^{ab}\eta^i_\pm = 0 \, .
\ee
The solutions to these equations with~$AdS_{2} \times S^{2}$ boundary conditions, the \emph{localizing manifold} of our problem, were found and analyzed in~\cite{Dabholkar:2010uh, Gupta:2012cy}. We rewrite the solutions in our conventions so as to have an easy reference for some calculations in Sections~\S\ref{formalism} and \S\ref{DetCalc}.

As explained in~\S\ref{formalism}, the metric and~$T$-tensor appearing in~\eqref{AppCKS1},~\eqref{AppCKS2} are given by~\eqref{metric},~\eqref{fieldconfattr} in the~$\sqrt{-g}=1$ gauge. We now observe that a set of solutions to the conformal Killing spinor equations can be found simply by setting
\be
\eta^i_\pm = 0 \, ,
\ee
taking~$\xi^i_\pm$ to be a solution of
\bea \label{AppKS}
2D_\mu\xi^i_+ + \tfrac{1}{16}T_{ab}^-\gamma^{ab}\gamma_\mu\xi^i_- &=& 0 \, , \\
2D_\mu\xi^i_- - \tfrac{1}{16}T^+_{ab}\gamma^{ab}\gamma_\mu\xi^i_+ &=& 0 \, , \nonumber
\eea
with~$AdS_2 \times S^2$ boundary conditions, and with the field~$D$ satisfying (with~$\xi^i = \xi^i_+ + \xi^i_-$)
\be \label{offshellBPSD}
\left(24D\gamma_5 + \gamma^\mu D_\mu T_{ab} \gamma^{ab}\right)\xi^i = 0 \, .
\ee
Note that in~\eqref{AppKS}, the covariant derivative only contains the spin-connection in our gauge-fixed theory 
since~$b_\mu$ and the R-symmetry gauge fields have been set to zero, 
i.e.,~$D_\mu\xi^i_\pm = \partial_\mu\xi^i_\pm - \tfrac{1}{4}\omega_\mu^{\;\;ab}\gamma_{ab}\xi^i_\pm$.

For our black hole solution in the gauge~$\sqrt{-g}=1$, we have
\be
T_{ab}^- = \begin{pmatrix} 0 & 4i & 0 & 0 \\ -4i & 0 & 0 & 0 \\ 0 & 0 & 0 & 4i \\ 0 & 0 & -4i & 0 \end{pmatrix}\, , \qquad T_{ab}^+ = \begin{pmatrix} 0 & 4i & 0 & 0 \\ -4i & 0 & 0 & 0 \\ 0 & 0 & 0 & -4i \\ 0 & 0 & 4i & 0 \end{pmatrix}\, ,
\ee
and~\eqref{AppKS} reduces to 
\be \label{AppKSfull}
D_\mu\xi^i = -\frac{1}{2}\left(\sigma_3\otimes1\right)\gamma_\mu\gamma_5\,\xi^i \, .
\ee
The solutions to~\eqref{AppKSfull} have been obtained for general~$AdS_n\times S^m$ geometries in~\cite{Lu:1998nu}. 
We parametrize the Euclidean~$AdS_2 \times S^2$ space as
\be \label{metricsameer}
ds^2 = g_{\mu\nu}dx^\mu dx^\nu = \sinh^2\eta \, d\tau^2 + d\eta^2 + d\psi^2 + \sin^2\psi \, d\phi^2 \, ,
\ee
to obtain the following four complex, linearly independent solutions of~\eqref{AppKSfull}:
\bea
\xi_1 &=& \sqrt{2}e^{\tfrac{i}{2}(\phi+\tau)} \begin{pmatrix} \sinh\tfrac{\eta}{2}\cos\tfrac{\psi}{2} \\ i\cosh\tfrac{\eta}{2}\cos\tfrac{\psi}{2} \\ -\sinh\tfrac{\eta}{2}\sin\tfrac{\psi}{2} \\ i\cosh\tfrac{\eta}{2}\sin\tfrac{\psi}{2} \end{pmatrix}\, , \quad
\xi_2 = \sqrt{2}e^{-\tfrac{i}{2}(\phi+\tau)} \begin{pmatrix} i\cosh\tfrac{\eta}{2}\sin\tfrac{\psi}{2} \\ -\sinh\tfrac{\eta}{2}\sin\tfrac{\psi}{2} \\ i\cosh\tfrac{\eta}{2}\cos\tfrac{\psi}{2} \\ \sinh\tfrac{\eta}{2}\cos\tfrac{\psi}{2} \end{pmatrix} \, , \\
\xi_3 &=& \sqrt{2}e^{\tfrac{i}{2}(\phi-\tau)} \begin{pmatrix} \cosh\tfrac{\eta}{2}\cos\tfrac{\psi}{2} \\ i\sinh\tfrac{\eta}{2}\cos\tfrac{\psi}{2} \\ -\cosh\tfrac{\eta}{2}\sin\tfrac{\psi}{2} \\ i\sinh\tfrac{\eta}{2}\sin\tfrac{\psi}{2} \end{pmatrix} \, , \quad
\xi_4 = \sqrt{2}e^{-\tfrac{i}{2}(\phi-\tau)} \begin{pmatrix} i\sinh\tfrac{\eta}{2}\sin\tfrac{\psi}{2} \\ -\cosh\tfrac{\eta}{2}\sin\tfrac{\psi}{2} \\ i\sinh\tfrac{\eta}{2}\cos\tfrac{\psi}{2} \\ \cosh\tfrac{\eta}{2}\cos\tfrac{\psi}{2} \end{pmatrix} \, . \nonumber
\eea
Spinors in Euclidean signature have a symplectic Majorana reality condition, and from the above complex solutions to~\eqref{AppKSfull}, one may generate symplectic Majorana solutions by taking the following combinations 
\bea \label{KSsol}
\xi^i_{(1)} &=& (\xi_1,-i\xi_2) \, , \quad \xi^i_{(2)} = (i\xi_1,-\xi_2) \, , \quad \xi^i_{(3)} = (\xi_2,i\xi_1) \, , \quad \xi^i_{(4)} = (i\xi_2,\xi_1) \, , \\
\xi^i_{(5)} &=& (\xi_3,-i\xi_4) \, , \quad \xi^i_{(6)} = (i\xi_3,-\xi_4) \, , \quad \xi^i_{(7)} = (\xi_4,i\xi_3) \, , \quad \xi^i_{(8)} = (i\xi_4,\xi_3) \, , \nonumber
\eea
where the~$SU(2)$ notation is~$\xi^i = (\xi^1,\xi^2)$. Moreover, the Weyl projection is compatible with the reality condition~\eqref{AppSM} and we may therefore build symplectic Majorana-Weyl solutions to~\eqref{AppKS} using the chirality projectors
\be
\xi^i_\pm = \frac{(1\pm\gamma_5)}{2}\xi^i\, . 
\ee
The procedure just described builds eight linearly independent, symplectic Majorana-Weyl solutions to~\eqref{AppCKS1},~\eqref{AppCKS2}, and these solutions all have~$\eta^i_\pm = 0$.

As explained in the introduction, in the context of localization we are interested in the supercharge which squares to~$(L_0 - J_0)$. This supercharge is parametrized by~$\xi^i_{(1)\,\pm}$ of~\eqref{KSsol}, whose Weyl-projected~$SU(2)$ components are explicitly given by
\bea \label{killingspinor}
\xi^1_{(1)\,+} &=& \frac{e^{-\tfrac{i}{2}(\tau + \phi)}}{\sqrt{2}}\begin{pmatrix} i\cosh\tfrac{\eta}{2}\sin\tfrac{\psi}{2} \\ 0 \\ 0 \\ \sinh\tfrac{\eta}{2}\cos\tfrac{\psi}{2} \end{pmatrix}\, , \quad
\xi^2_{(1)\,+} = \frac{e^{\tfrac{i}{2}(\tau + \phi)}}{\sqrt{2}}\begin{pmatrix} i\sinh\tfrac{\eta}{2}\cos\tfrac{\psi}{2} \\ 0 \\ 0 \\ -\cosh\tfrac{\eta}{2}\sin\tfrac{\psi}{2} \end{pmatrix}\, , \\
\xi^1_{(1)\,-} &=& \frac{e^{-\tfrac{i}{2}(\tau + \phi)}}{\sqrt{2}}\begin{pmatrix} 0 \\ -\sinh\tfrac{\eta}{2}\sin\tfrac{\psi}{2} \\ i\cosh\tfrac{\eta}{2}\cos\tfrac{\psi}{2} \\ 0 \end{pmatrix}\, , \quad \!\!\!
\xi^2_{(1)\,-} = \frac{e^{\tfrac{i}{2}(\tau + \phi)}}{\sqrt{2}}\begin{pmatrix} 0 \\ -\cosh\tfrac{\eta}{2}\cos\tfrac{\psi}{2} \\ -i\sinh\tfrac{\eta}{2}\sin\tfrac{\psi}{2} \\ 0 \end{pmatrix}\, . \nonumber
\eea

The supercharge~$Q$ built out of this spinor satisfies the algebra~\eqref{Qsquarefullcov}. The fermionic bilinears
\be
v^\mu = -2i\epsilon_{ij} \, \bar{\xi}^i_+ \gamma^\mu\,\xi^j_-\, , \qquad  \!\!\!\!K_\pm = \epsilon_{ij}\,\bar{\xi}_\pm^i \xi_\pm^j \, , 
\ee
that are used in the main text, have the following expressions:
\be \label{vmu1}
v^\mu \=  \bigl( -1 \quad  0 \quad  0 \quad  1 \bigr)^{T} \, , 
\ee
and 
\be \label{kpm1}
K_\pm \= \frac{1}{2}\left(\pm\cos\psi - \cosh\eta\right) \, .
\ee

\section{Transversally elliptic operators and the symbol of~$D_{10}$ \label{symbol}}

The standard starting point for the considerations of index theory is that of an \emph{elliptic operator} on a manifold, 
which generalizes the notion of a Laplacian. If the operator is linear and of second order, 
we can write it in local coordinates~$x^{i}$ as 
\be
a^{ij} (x) \, \p_{i} \, \p_{j} + b^{i} (x) \p_{i} + c_{i} (x) \, . 
\ee
An elliptic operator is one for which the matrix~$a^{ij}$ is positive-definite\footnote{For technical reasons, the 
theory of elliptic operators often also assumes that the eigenvalues are bounded.}. This can be  restated 
as follows: if we replace the derivatives by momenta, i.e.~consider the Fourier transform of the linear operator, 
we obtain the \emph{symbol} of the operator. 
An operator is elliptic if the \emph{principal symbol}~$a^{ij} \, p_{i} \, p_{j}$ does not vanish for any non-zero~$p_{i}$.

Our operator~$D_{10}$, however, is not elliptic -- but it can still be 
treated by index theory~\cite{Atiyah:1974}. 
The point is that we have a certain special~$U(1)$ action (that of~$H$), and our operator~$D_{10}$ 
commutes with this action. 
In the directions transverse to the~$U(1)$ orbits, the operator~$D_{10}$ \emph{is} elliptic -- such an operator 
is called \emph{transversally elliptic}, and there is a version of the index theorem that deals with such operators. 
In terms of the symbol, an operator is called transversally elliptic if its symbol does not vanish for any~$p_{i}$ that 
is transversal to the generator of the~$U(1)$ action. 
This means that the matrix~$a_{ij}$ is allowed to degenerate, but only along the 
one-dimensional locus generated by the~$U(1)$ action. 
We show below that the operator~$D_{10}$ is transversally elliptic with respect to the~$U(1)$ symmetry 
generated by~$H$.

As explained in section~\S\ref{DetCalc}, the one-loop determinant acquires contributions from vector and hyper 
multiplets separately. We therefore need to prove that the operators~$D_{10}^{\text{vect}}$ in the vector multiplet sector and~$D_{10}^{\text{hyp}}$ in the hyper multiplet sector are both transversally elliptic with respect to the~$U(1)$ action generated by~$H$.

\subsection*{Vector multiplets}

To read off the form of the operator~$D^{\text{vect}}_{10}$, we introduce the following quantities:
\bea \label{diagonalvars}
\Lambda^I &\equiv& Q(X^I - \bar{X}^I) = \epsilon_{ij}\left(\bar{\xi}^i_+\lambda^{I\;j}_+ - \bar{\xi}^i_-\lambda^{I\;j}_-\right)\, , \cr
\Lambda^I_\mu &\equiv& QA_\mu^I = \epsilon_{ij}\left(\bar{\xi}^i_-\gamma_\mu\,\lambda^{I\;j}_+ - \bar{\xi}^i_+\gamma_\mu\,\lambda^{I\;j}_-\right)\, , \\
\Xi^{I\;ij} &\equiv& 2\left(\bar{\xi}^{(i}_+\lambda^{I\;j)}_+ + \bar{\xi}^{(i}_-\lambda^{I\;j)}_-\right) \, .\nonumber
\eea
We split the fields of the vector multiplet (including the BRST ghosts) into bosons and fermions according to
\be \label{fieldsplit1}
X = \left\{A_\mu^I,X^I-\bar{X}^I \right\} \, , \qquad \Psi = \left\{\Xi^{I\;ij},b^I,c^I\right\} \, ,
\ee
and their~$\wh{Q}$-superpartners
\be \label{fieldsplit2}
\wh{Q}\Psi = \left\{\wh{Q}\Xi^{I\;ij},B^I,-\wh{\theta}^I\right\} \, , \qquad \wh{Q}X = \left\{\wh{Q}A_\mu,\wh{Q}(X^I-\bar{X}^I)\right\} \, .
\ee
The relations~\eqref{diagonalvars} may be inverted to yield
\bea
\lambda^{I\;i}_+ &=& \frac{-1}{\cosh\eta}\left(\xi^i_+\Lambda^I + \gamma^\mu\xi^i_-\Lambda_\mu^I - \epsilon_{kl}\xi^k_+\Xi^{I\;il}\right) \, , \\
\lambda^{I\;i}_- &=& \frac{1}{\cosh\eta}\left(\xi^i_-\Lambda^I + \gamma^\mu\xi^i_+\Lambda_\mu^I + \epsilon_{kl}\xi^k_-\Xi^{I\;il}\right) \, .
\eea

The deformation operator in the vector multiplet sector,~$\wh\CV^{\,\text{vect}}$, is as follows:
\be
\wh\CV^{\,\text{vect}} = \int d^4x \sum_I \left[\left(\wh{Q}\lambda^i_+\, , \,\lambda^{I\;i}_+\right) + \left(\wh{Q}\lambda^i_-\, , \,\lambda^{I\;i}_-\right) + b^I G^A(A^I_\mu)\right]\, ,
\ee
As explained in App.~\ref{euclideanspinors}, we will take the inner product in the equation above to be the hermitian 
conjugate in Minkowski signature, compute the quantity in the right-hand side 
using this inner product and the original Minkowski reality conditions on the fields, and only conduct the Euclidean 
continuation at the very end by imposing different reality conditions on the various fields. Thus, we write:
\be
\wh\CV^{\,\text{vect}} = \int d^4x \sum_I \left[\left(\wh{Q}\lambda^i_+\right)^\dagger\,\lambda^{I\;i}_+ + \left(\wh{Q}\lambda^i_-\right)^\dagger\,\lambda^{I\;i}_- + b^I G^A(A^I_\mu)\right]\, ,
\ee
In terms of the variables~\eqref{diagonalvars}, this is
\be
\wh\CV^{\,\text{vect}} = \int d^4x \sum_I \Big[\frac{1}{\cosh\eta}\left(\left(\wh{Q}\Lambda^I\right)^\dagger\Lambda^I + \left(\wh{Q}\Lambda_\mu^I\right)^\dagger\Lambda^{I\,\mu} + \frac{1}{2}\left(\wh{Q}\Xi^{I\,ij}\right)^\dagger\Xi^{I\,ij}\right) + b^I G^A(A^I_\mu)\Big] \, .
\ee

To compare to the general expression for the quadratic fluctuations~\eqref{deformationoperator}, we express the regulator in terms of the fields~\eqref{fieldsplit1},~\eqref{fieldsplit2}. We write the various terms in the equation above in terms of these fields by making use of the following relations:
\bea
\Lambda^I &=& \wh{Q}(X^I - \bar{X}^I)\, , \;\; \left(\wh{Q}\Lambda^I\right)^\dagger = iv^\mu\partial_\mu(X^I - \bar{X}^I) \, , \\
\Lambda^I_\mu &=& \wh{Q}A^I_\mu - \partial_\mu c^I\, , \;\;\; \left(\wh{Q}\Lambda^I_\mu\right)^\dagger = -iv^\nu F^I_{\nu\mu} + iv^\nu\partial_\mu A^I_\nu + 2\partial_\mu\left(\cos\psi(X^I-\bar{X}^I)\right) - \partial_\mu\,\wh{Q}c^I \, . \nonumber
\eea
Here 
we use the Minkowksi reality conditions on the vector multiplet fields:
\be
(X^I)^\dagger = \bar{X}^I \, \qquad (A^I_\mu)^\dagger = A^I_\mu \, ,
\ee
as consistent with the original~$\CN=2$ Minkowski superconformal algebra.

Further, we have:
\be
\wh{Q}\Xi^{I\;ij} = \left(\bar{\xi}^{(i}_+\gamma^{ab}\xi^{j)}_+ + \bar{\xi}^{(i}_-\gamma^{ab}\xi^{j)}_-\right)\mathcal{F}^I_{ab} + 4\,\bar{\xi}^{(i}_+\gamma^\mu\xi^{j)}_-\,\partial_\mu(X^I - \bar{X}^I) + 2\left(\bar{\xi}^k_+\xi^{(i}_+ + \bar{\xi}^k_-\xi^{(i}_-\right)Y_k^{I\;j)} \, ,
\ee
where~$\mathcal{F}_{ab}^I = F_{ab}^I - \tfrac{1}{4}\left(\bar{X}^I T_{ab}^- + X^I T_{ab}^+\right)$, and we can use this equation to express the auxiliary~$Y^I_{ij}$ in terms of the cohomological fields as follows:
\be \label{Ycohom}
2\left(\bar{\xi}^k_+\xi^{(j}_+ + \bar{\xi}^k_-\xi^{(j}_-\right)Y_k^{I\;l)} = \wh{Q}\Xi^{I\;jl} - \left(\bar{\xi}^{(j}_+\gamma^{ab}\xi^{l)}_+ + \bar{\xi}^{(j}_-\gamma^{ab}\xi^{l)}_-\right)\mathcal{F}^I_{ab} - 4\,\bar{\xi}^{(j}_+\gamma^\mu\xi^{l)}_-\,\partial_\mu(X^I - \bar{X}^I) \, .
\ee 
From this we deduce
\bea
\left(\wh{Q}\Xi^{I\;ij}\right)^\dagger &=& \epsilon_{ik}\epsilon_{jl}\left[\wh{Q}\Xi^{I\;kl} - 2\left(\bar{\xi}^{(k}_+\gamma^{ab}\xi^{l)}_+ + \bar{\xi}^{(k}_-\gamma^{ab}\xi^{l)}_-\right)F^I_{ab} - 8\,\bar{\xi}^{(k}_+\gamma^\mu\xi^{l)}_-\,\partial_\mu(X^I - \bar{X}^I)\right] \cr
&+&\tfrac{1}{4}\epsilon_{ik}\epsilon_{jl}\left(\bar{\xi}^{(k}_+\gamma^{ab}\xi^{l)}_+ + \bar{\xi}^{(k}_-\gamma^{ab}\xi^{l)}_-\right)(T^+_{ab}-T^-_{ab})(X^I-\bar{X}^I) \, .
\eea
Once these expressions have been derived, we can analytically continue to Euclidean by imposing 
the real slice in which~$X$ and~$\bar{X}$ are independent real variables. 

Collecting all this, we can write the terms of the regulator~$\wh\CV^{\,\text{vect}}$ relevant for the symbol computation in each vector multiplet~$I$:
\bea
&&i\,\partial^\mu c^I v^\nu F^I_{\nu\mu} - i\,\partial^\mu c^I v^\nu\partial_\mu A^I_\nu - 2\,\partial^\mu c^I\partial_\mu\left(\cos\psi(X^I-\bar{X}^I)\right) + b^I G^A(A^I_\mu)  \\
&&- \epsilon_{ik}\epsilon_{jl}\,\Xi^{I\,ij}\left(\bar{\xi}^{(k}_+\gamma^{ab}\xi^{l)}_+ + \bar{\xi}^{(k}_-\gamma^{ab}\xi^{l)}_-\right)F^I_{ab} - 4\epsilon_{ik}\epsilon_{jl}\,\Xi^{I\,ij}\,\bar{\xi}^{(k}_-\gamma^\mu\xi^{l)}_+\,\partial_\mu(X^I - \bar{X}^I) \, . \nonumber
\eea
We are interested in the symbol of the~$D^{\text{vect}}_{10}$ operator,~$\sigma\left(D^{\text{vect}}_{10}\right)$, which is obtained by 
replacing derivatives according to~$\partial_a \rightarrow i p_a$, where~$p_a$ are momenta. In order for the 
index theorem to apply, we wish to verify that~$D^{\text{vect}}_{10}$ is transversally elliptic with respect
to the~$U(1)$ symmetry generated by~$H$. Therefore, we want to check that the determinant of its symbol is non-zero 
everywhere in~$AdS_2 \times S^2$ as long as momenta transverse to the direction specified by~$v^\mu$ are turned on. 
We still have some freedom in choosing the gauge-fixing function~$G^A$. If we take the usual Lorentz gauge
\be
G(A_\mu^I) = \partial^\mu A_\mu^I \, ,
\ee 
we find that the determinant of the symbol in each vector multiplet~$I$ is given by
\be
\text{det}\left[\sigma\left(D^{\text{vect}}_{10}\right)\right] = -2\,(p_i^2+p_v^2)\left(p_i^2 \cosh^2\eta + p_v^2 \cos^2\psi\right)^2 \, ,
\ee 
where~$p_i$,~$i=1,2,3$, are momenta transverse to the~$H$ action and~$p_v$ is the momenta along~$v^\mu$. We clearly see that, 
when~$p_i = 0$, the determinant vanishes on the equator of the~$S^2$. When some transverse momentum is turned on, this degeneracy 
is lifted and the determinant of the symbol is non-zero everywhere in space-time. This shows that the operator~$D^{\text{vect}}_{10}$ 
for each vector multiplet~$I$ is indeed transversally elliptic. 

\subsection*{Hyper multiplets}

In the hyper multiplet sector, we introduce
\bea \label{diagonalvarshyper}
\Lambda_i^{\;\alpha} &\equiv& QA_i^{\;\alpha} = 2\,\epsilon_{ij}\left(\bar{\xi}^j_-\zeta^\alpha_- - \bar{\xi}^j_+\zeta^\alpha_+\right)\, , \cr
\Xi_i^{\;\alpha} &\equiv& 2\,\epsilon_{ij}\left(\bar{\breve{\xi}}^{\,j}_+\zeta^\alpha_+ + \bar{\breve{\xi}}^{\,j}_-\zeta^\alpha_-\right)\, .
\eea
We split the fields according to
\be
X = \left\{A_i^{\;\alpha}\right\} \, , \qquad \Psi = \left\{\Xi_i^{\;\alpha}\right\} \, ,
\ee
and their~$\wh{Q}$-superpartners. We can once again invert the relations~\eqref{diagonalvarshyper} to write
\be
\zeta^\alpha_+ = \frac{-1}{\cosh\eta}\left(\Lambda_i^{\;\alpha}\xi^i_+ - \Xi_i^{\;\alpha}\breve{\xi}^{\,i}_+\right)\, , \quad \zeta^\alpha_- = \frac{1}{\cosh\eta}\left(\Lambda_i^{\;\alpha}\xi^i_- + \Xi_i^{\;\alpha}\breve{\xi}^{\,i}_-\right) \, .
\ee
The deformation operator for one hyper multiplet is now
\be
\wh\CV^{\,\text{hyp}} = \int d^4x \left[\left(\wh{Q}\zeta^\alpha_+\right)^\dagger\zeta^\alpha_+ + \left(\wh{Q}\zeta^\alpha_-\right)^\dagger\zeta^\alpha_-\right]\, ,
\ee
In terms of the variables~\eqref{diagonalvarshyper}, this is
\be
\wh\CV^{\,\text{hyp}} = \int d^4x \, \frac{1}{2\cosh\eta}\left[\left(\wh{Q}\Lambda_i^{\;\alpha}\right)^\dagger\Lambda_i^{\;\alpha} + \left(\wh{Q}\Xi_i^{\;\alpha}\right)^\dagger\Xi_i^{\;\alpha}\right] \, .
\ee
We have
\be
\left(\wh{Q}\Xi_i^{\;\alpha}\right)^\dagger = \wh{Q}\Xi^i_{\;\alpha} + 4\,\rho_{\alpha\beta}\left(\bar{\breve{\xi}}^{\;i}_+\gamma^\mu\xi^j_- + \bar{\breve{\xi}}^{\;i}_-\gamma^\mu\xi^j_+\right)\partial_\mu A_j^{\;\beta} \, ,
\ee
where we've again made use of the Minkowski reality condition
\be
\left(A_i^{\;\alpha}\right)^\dagger = \epsilon^{ij}\,\rho_{\alpha\beta}\,A_j^{\;\beta} \, .
\ee
We see that the relevant term for the symbol of~$D_{10}^{\text{hyp}}$ is simply
\be
4\,\rho_{\alpha\beta}\,\Xi_i^{\;\alpha} \left(\bar{\breve{\xi}}^{\;i}_+\gamma^\mu\xi^j_- + \bar{\breve{\xi}}^{\;i}_-\gamma^\mu\xi^j_+\right)\partial_\mu A_j^{\;\beta} \, .
\ee
To compute the symbol and its determinant, we therefore need an explicit solution to the constraint equations~\eqref{constrainedparam}. We choose:
\be \label{xicheck}
\breve{\xi}^i_+ = \left(\frac{\cosh\eta-\cos\psi}{\cosh\eta+\cos\psi}\right)^{-1/2}\xi^i_+ \, , \qquad \breve{\xi}^i_- = \left(\frac{\cosh\eta-\cos\psi}{\cosh\eta+\cos\psi}\right)^{1/2}\xi^i_- \, ,
\ee
and we find for the determinant of the symbol of~$D_{10}^{\text{hyp}}$
\be
\text{det}\left[\sigma\left(D^{\text{hyp}}_{10}\right)\right] = 2\left(p_i^2\cosh^2\eta + p_v^2\cos^2\psi\right)\, , 
\ee
where we use the momenta and notation introduced for the vector multiplet case. We again see that the symbol is non-invertible along the equator of~$S^2$ when only the momentum along the~$H$ action is non-zero, and this degeneracy is lifted whenever some transverse momenta are turned on. Thus the operator~$D^{\text{hyp}}_{10}$ is transversally elliptic with respect to the~$H$ action.

\bibliography{AdS2dets}{}
\bibliographystyle{JHEP}

\end{document}